\documentclass{aastex61}
\usepackage[flushleft]{threeparttable}
\usepackage{color}
\usepackage{float}
\usepackage[caption = false]{subfig}

\newcommand\aastex{AAS\TeX}

\newcommand{\NII}{[\ion{N}{2}]}
\newcommand{\OIII}{[\ion{O}{3}]}

\submitjournal{ApJ}

\shorttitle{\aastex\ AGN identification on stellar population models}
\shortauthors{Chen et al.}

\def\kms{\mbox{${\rm km}\,{\rm s}^{-1}$}}  

\begin{document}


\title{Dependence of optical  Active Galactic Nuclei identification \\ on stellar population models}

\correspondingauthor{Yan-Ping Chen}
\email{yc47@nyu.edu}

\author{Yan-Ping Chen}
\affil{New York University Abu Dhabi, P.O. Box 129188,\\
 Abu Dhabi, United Arab Emirates}

\author{Ingyin Zaw}
\affiliation{New York University Abu Dhabi, P.O. Box 129188,\\
 Abu Dhabi, United Arab Emirates}
\affiliation{Center for Cosmology and Particle Physics, \\Physics Department, New York University, \\New York, NY 10003, USA.}
 

\author{Glennys R Farrar}
\affiliation{Center for Cosmology and Particle Physics, \\Physics Department, New York University, \\New York, NY 10003, USA.}






\begin{abstract}

We have conducted a study to quantify the systematic differences resulting from using different stellar 
population models in optical spectroscopic identification of type II AGN. We examined the different 
AGN detection fractions of 7069 nearby galaxies ($z\le0.09$) with SDSS DR8 spectra when using the
\citet[][BC03]{BC03}, \citet[][MILES]{Vazdekis10}, and solar metallicity \citet[][]{Maraston11} 
(MS11$_{solar}$) stellar population models. 
The line fluxes obtained using BC03 and MS11$_{solar}$ are publicly available from SDSS data releases.
We find that the BC03 templates result in systematically higher BPT line ratios and consequently higher AGN fractions and the 
MS11$_{solar}$ templates result in systematically lower line ratios and AGN fractions compared with the MILES templates. 
Using MILES as the standard, BC03 results in 25\% ``false positives'' and MS11$_{solar}$ results in 22\% ``false negatives''
when using the \citet{K01} boundary for AGN identification.
The fraction of galaxies whose AGN identification changes for different templates is luminosity dependent, ranging from a few percent for L$_{[O III]5007}$  $ \ge 10^{40}$ erg $s^{-1}$ and increasing to $\sim 50 \%$ for L$_{[O III]5007}$  $ \le 10^{38}$ erg $s^{-1}$.
These results suggest that template choice should be accounted for when using and comparing the AGN and emission
 line fluxes from different catalogs. 
 
\end{abstract}




\section{Introduction} \label{sec:intro}

Optical spectroscopy is a powerful tool to probe the physical properties of active galactic nuclei (AGN). 
 Broad-line (type I) AGN are characterised by their broad hydrogen Balmer lines, with Full Width at Half Maximum
 (FWHM) values up to a few thousand $\kms$. Narrow-line (type II) AGN however, need to be distinguished 
 from star-forming galaxies since they emit the same set of  ionised forbidden lines such as \NII\ 6584
 and \OIII 5007. The differences in the ratios of these lines to the narrow hydrogen Balmer lines (i.e., H${\alpha}$, 
 and H${\beta}$) between AGN and star-forming galaxies were first reported by \citet[][hereafter, BPT]{bptref},
and further explored by others \citep[e.g.,][]{Osterbrock85, Veilluex87, K01, K03, Stasinska06, Kewley06,
Cidfernandes10, Cidfernandes11}.
 \citet{K01} developed separation criteria based on theoretical modeling of star formation lines, 
while \citet{K03} defined empirical separation criteria based on Sloan Digital sky Survey \citep[SDSS,][]{York00} spectra.

The spectra of galaxies containing AGN show
not only emission features but also stellar absorption lines
and continuum emission from the host galaxies.
Subtraction of the host galaxy contribution to the integrated light
in previous works \citep[e.g.,][]{Ho97,Kewley00, Kewley01b, K03} has
been done either with a local polynomial fit 
or using templates \citep[e.g.,][]{BC03}. 
The former is easy to perform, does not rely on any models, and can be straightforwardly applied as a 
quick solution for the absorption and continuum components estimation, especially for objects with 
very strong emission features (e.g., those of Seyfert I AGN). However, when the spectrum contains 
 non-negligible absorption features, the accuracy of the emission line fluxes can be affected. 
 Weak emission line features, especially hidden in the absorption and 
continuum components, can easily be underestimated in a local polynomial continuum fit. In these cases,
using stellar templates  to fit the absorption and continuum gives a more accurate solution.
Full-spectrum-fitting and subtraction of absorption and continuum components requires strong enough 
absorption features to match with stellar templates. Instead of full-spectrum-fitting,
one can also use principle component analysis (PCA) \citep[e.g.,][]{Hao05a, Greene07, Allen13} 
to determine the underlying galaxy spectra and to extract the emission lines.

Several studies based on AGN spectroscopic identification have
 been published \citep[e.g.,][]{K03, Miller03,Kewley06} using SDSS data. 
The AGN detection rate is known to depend on the details of data processing \citep{Hao05a}: redshift ranges,
signal-to-noise cut of the data, and the boundary that separates type II AGN from star-forming galaxies
 in BPT diagrams (i.e., the authors may have different AGN identification criteria).

In this work, 
we consider the impact of the using different stellar population models on AGN identification, which has not previously been quantified. 
The AGN identification in SDSS data releases first used \citet[][BC03]{BC03}, and currently 
\citet[][]{Maraston11} at solar metallicity only (MS11$_{solar}$), as templates for
 stellar component analysis.
As more stellar population models become available, systematic studies of AGN identification's dependence on the stellar model used
become possible. This is important not only for finding the best template, but also for understanding the merits and limitations of the models.  
In our study, we compare the results from BC03, MS11$_{solar}$, and \citet[][MILES]{Vazdekis10} stellar population
models.
  
  In Section~\ref{secdata}, we present both the 
sample for our study and our procedure for identifying narrow line AGN candidates. 
In Section~\ref{secmodel}, we review the stellar population models used in this analysis.
 We also describe how we applied stellar population models to our data. Our results and  the relative effects from varing 
 the templates are presented in Section~\ref{sectest}, where we also discuss the aspects of the stellar templates 
 which lead to different results. We present our conclusions in Section~\ref{secconclusion}. 
The appendix includes other properties we explored that do not cause major differences in 
type II AGN identification, namely, wavelength range, young stellar populations, Seyfert II vs. LINERs, and data quality.


\section{Data sample and AGN identification}\label{secdata}

This work is part of our efforts to construct an all-sky optical AGN catalog  (Zaw, Chen and Farrar, in prep, ZCF18 below),
based on optical spectra from the 2MASS Redshift Survey \citep[2MRS,][]{Huchra12}.
The 2MRS was assembled from observations by the 2MRS team (with the FAst Spectrograph for the Tillinghast telescope at the 
Fred L. Whipple Observatory in the north and Cerro Tololo Interamerican Observatory in the south) and from other catalogs, including 
the SDSS data release (DR) 8, the 6dF Galaxy Survey, and the NASA Extragalactic Database.
Among all the subsamples, the SDSS subsample has the best signal to noise ratios, and
the only one where absolute flux calibration and telluric correction has been applied to the spectra.
In addition, the line fluxes from SDSS data releases can be used to cross check our work.
We therefore use the SDSS subsample for our study.
The SDSS subsample consists of the spectra of 7069 galaxies with a redshift of $z \leq 0.09$.
These spectra cover the wavelength range 3800--9200 \AA, with a mean resolution of 
R$\sim$1800--2000. 

\begin{figure}[ht!]
\plotone{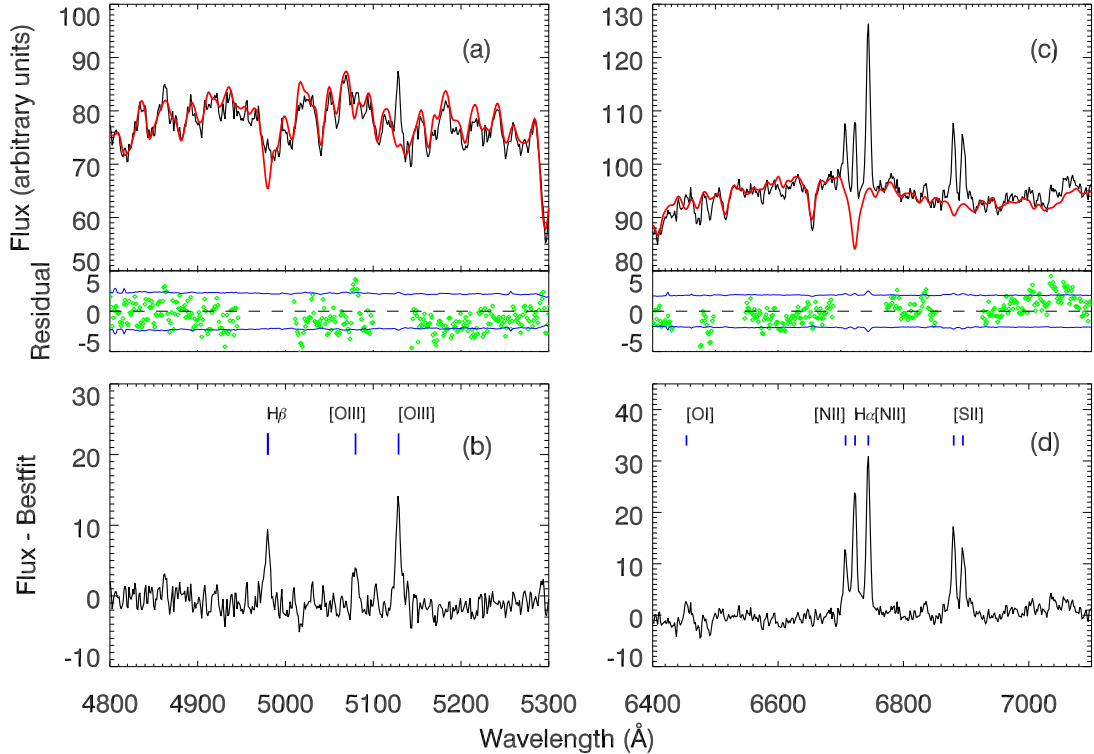}
\caption{An example to illustrate that template subtraction is necessary to identify weak emission lines. Top panels: data are marked as black lines, the best fit models are marked as red lines, the horizontal
blue lines are showing the 1$\sigma$ error spectra, the green dots are the residuals in regions without
emission lines. Weak H${\beta}$ emission line is absent before template subtraction. Bottom panels: Residual spectra showing the emission line features after subtracting the best-fit (absorption) models.  }
\label{fig:weakhb}
\end{figure}

\subsection{Type II AGN candidate identification}

A detailed description of the emission
line galaxy selection is given in our catalog paper (ZCF18). 
We briefly summarise the process here. 
After rebinning the data and model spectra to the same spectral resolution, and masking out the AGN 
identification emission line regions, a full-spectrum-fitting was performed on each SDSS object using 
pPXF \citep{Cappellari04}. 
In the fitting procedure, the model spectra are shifted to the redshift of 
the observed spectra and broadened to account for stellar velocity dispersion. 
Each fit consists of a linear combination of model spectra.
The fits are required to yield a physical stellar velocity dispersion, $\sigma_{\rm fit} < 1000\ \kms $, 
in order to be considered acceptable. The fitting routine uses the error spectrum, provided with the data, to calculate 
a reduced $\chi^2$ value for each acceptable fit.
The ``best-fit'' spectrum is the one which
minimizes the reduced $\chi^2$. 
We limit the sample to those spectra whose reduced ${\chi^2}$ are less 
than 2.55 in the full-spectrum-fitting process, which keeps $99\%$ of the spectra with successful fits. 

The residual spectrum produced by subtracting the best-fit model from the data spectrum, is used to
analyze the emission line features. The galaxies with weak emission lines are 
most affected by the choice of stellar templates.
Figure~\ref{fig:weakhb} shows an example where the H${\beta}$ 
emission line is invisible in the spectrum by eye, but is identified by the template subtraction
using MILES \citep{Vazdekis10} templates.  

In this work, we use  $\rm{H{\alpha}}$,
 $ \rm{H{\beta}}$,  \OIII\ 5007, and \NII\ 6584 to identify galaxies as type II AGN. 
We infer the flux 
for each of the emission lines from the residual spectrum, by fitting Gaussian profiles. 
The line fluxes are calculated under the fitted Gaussian profiles within {3$\sigma$ of the line peak, where $\sigma$
is the fitted Gaussian width of the line.
The flux errors (i.e. line noises) are calculated as the sum in quadrature of the error spectra under the 
same wavelength range of the emission lines.
 An emission line is defined as one where the line flux divided by the flux error  
 is greater than three (i.e. S/N$\ge$3) when using MILES templates. 
We define emission line galaxies as those where all four diagnostic 
lines ($\rm {H}{\beta}$, \OIII\ 5007, $\rm {H}{\alpha}$ and \NII\ 6584) have S/N$\ge$3. 
There are 3350 emission line galaxies in our sample.
We use the BPT line ratios \NII/$\rm H{\alpha}$ and \OIII/$\rm H{\beta}$ and the \citet{K01} criteria 
to separate 
star-forming galaxies and type II AGN candidates, i.e., those above \citet{K01} line are taken as type II AGN,
those below \citet{K01} line are taken as star-forming galaxies. 

\section{Model Templates}\label{secmodel}

Stellar population models (SPM) are constructed by integrating a group of stellar 
spectra, known as a stellar library, with weights given by the initial mass functions. 
The group of stars are assumed to share the 
same age and chemical components, (i.e., metallicity), but with a variety of stellar masses.
Their integrated light forms a single stellar population spectrum. 
The connection between stellar population models and individual stellar spectra are stellar parameters
such as effective temperature $T_{\mathrm{eff}}$, surface gravity $\log g$, and metallicity $\rm [Fe/H]$. 
 When the stellar populations are constructed, an interpolator is applied to generate the grid of stars
 used to integrate the light of the population; see
 \citet{Conroy13rev}, \citet{Vazdekis10} and references therein for details.
The interpolator is based on the parameter space coverage of the underlying stellar library.
When stars are limited to a certain part of the parameter space, the interpolator may be biased if the available 
data does not adequately span the parameter space. 
Each stellar population model consists of a set of single stellar population (SSP) templates with
a given age and metallicity.

The \citet{BC03} templates are widely used to subtract the stellar absorption 
components and the continuum, in analyses of emission line galaxies, because they cover 
a large spectral wavelength range and were developed earlier.
In the last decade, with improvements in both data and theoretical
modelling, more stellar libraries have been constructed and published \citep[e.g., ][]{XSL, milesref, ngsl, elodie}, 
and their corresponding stellar population models are available 
\citep[e.g.,][]{Vazdekis10, pegase, ngslmodel, Maraston11}.

In this work we compare the 
stellar population models of MILES,  MS11$_{solar}$ MILES-based models,
\citet[][hereafter G05]{G05} and BC03.
The SDSS fluxes for DR8 were
 based on host galaxy subtraction using BC03.  More recently, \citet{Thomas13} made public new fluxes for 
 DR8 spectra using MS11$_{solar}$. MILES model is based on observed stellar 
optical spectra \citep{milesref}; the BC03 model was constructed from a combination of theoretical 
and observed spectra. The MS11$_{solar}$ MILES  based model shares the same input stellar library as MILES model.
The G05 model was constructed from theoretical stellar spectra, and is the best available theoretical model.
We use them when empirical data are incomplete, e.g., for young stellar populations. 
The model ingredients are described in section~\ref{modelingre}.
Since the MILES models are built from the MILES stellar library, currently the best empirical optical stellar library
and widely used by several stellar population models (e.g., \cite{Conroy10,Maraston11,Vazdekis10,MIUSCAT}),
we therefore use the MILES templates as our standard. 

\begin{figure} 
\includegraphics[scale=0.7,angle=90]{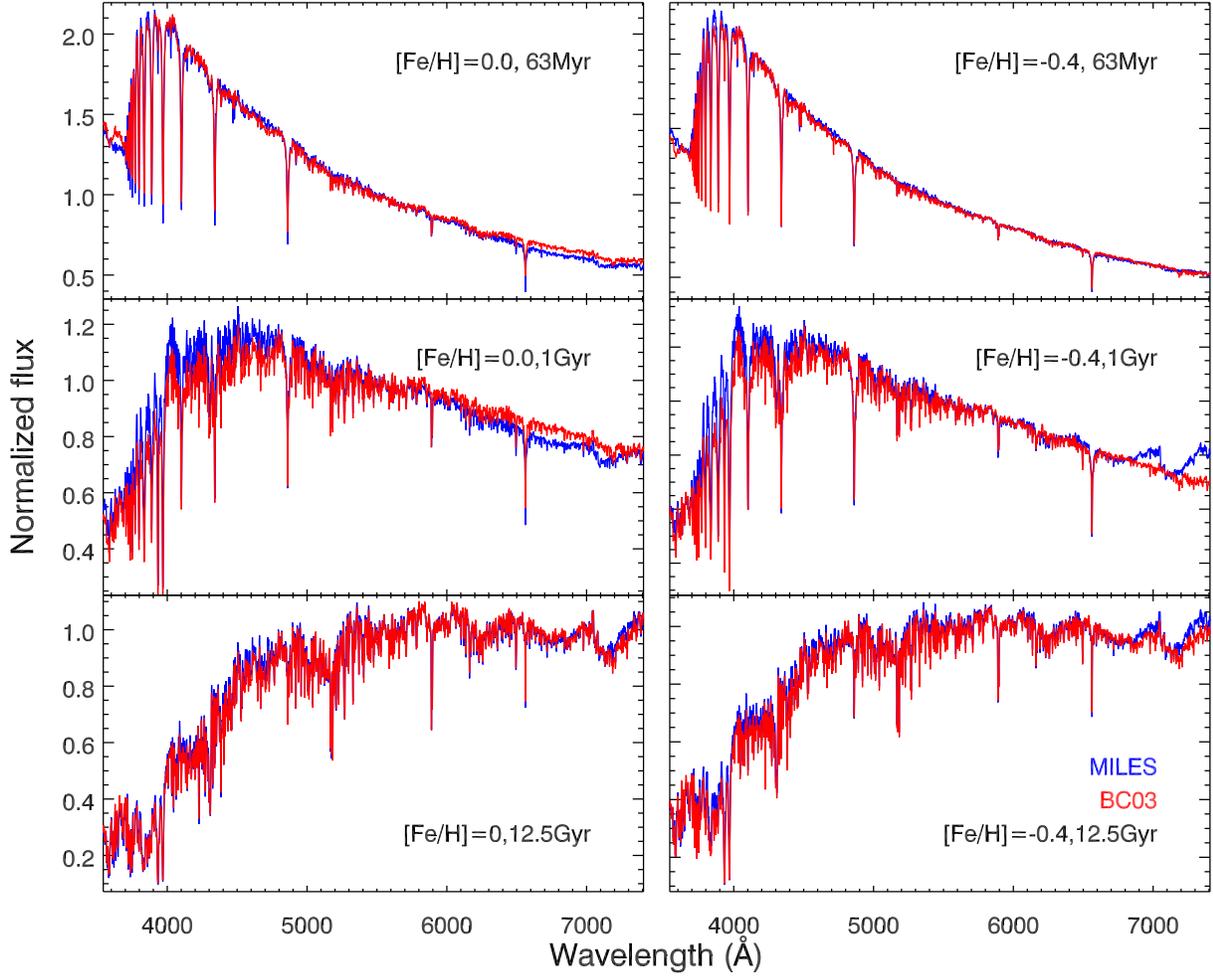}
\caption{Full spectrum model comparison, spectra were normalised at $\lambda \lambda$ 5500\AA. MILES 
 templates were smoothed to the same resolution as BC03 for fair comparison. Note the discrepancies at long wavelength, and the systematically lower flux of BC03 in the 4000--5000 \AA\ range in the middle panels. }
\label{fig:milesbc03full}
\end{figure}

\subsection{Differences of stellar population models}\label{modelingre}

Stellar population models have several distinct properties. 
The parameter ranges for the models we compare in this work are given in Table \ref{Sspmodels}.
They vary in age and metallicity ranges, as well as in stellar libraries. We will investigate how each of these affects our results. 
The key ingredients of stellar population models that describe the individual spectra of stellar systems
from different sources are as follows. BC03 mainly uses a combination of the empirical stellar library
STELIB \citep{stelibref} 
and a series of theoretical stellar libraries BaSeL \citep{lejeune97, lejeune98,Westera02}. 
The BC03 SSPs  are likely to be biased in the optical, having been built from the STELIB library that contains 
249 stars, with very few stars at non-solar metallicities. 
The MILES model is based on the empirical MILES stellar library that covers
a wide range of stellar parameter space (i.e., $T_{\mathrm{eff}}$, $\log g$, and $\rm [Fe/H]$), with
four times more stars than in STELIB. 
The G05 model is based on a theoretical stellar library; 
they contain a wide age range, helpful in understanding 
the contribution from younger stellar components in the host galaxies.
The MILES-based MS11 model used the same stellar library as MILES, with a special procedure for
integrating the light of thermally pulsing asymptotic giant branch (TP-AGB) stars. 
The MS11$_{solar}$ templates are the subset of MILES-based MS11 SPM that assume solar metallicity
abundances.

\begin{table}[ht!]
\centering
\caption{Stellar population models used for this work and their properties.}
 \begin{threeparttable}\label{Sspmodels}
 \footnotesize 
\begin{tabular}{|l|l|l|l|l|}

\hline
  \multicolumn{1}{|c|}{Source} &
  \multicolumn{1}{c|}{Wavelength range} &
  \multicolumn{1}{c|}{Resolution} &
  \multicolumn{1}{c|}{Age range} &
  \multicolumn{1}{c|}{Metallicity (Z) range } \\
\hline
  BC03 &    91\AA-160$\mu$m & 3\AA   & 0.1Myr-20Gyr & Z$=$0.004 -- 0.05 \\
  MILES ssp &   3500-7500\AA & 2.51\AA & 0.06-15Gyr & Z$=$0.0001 -- 0.03 \\
  G05 &    3000-7000\AA & 0.3\AA & 4Myr-17Gyr & Z$=$0.004 -- 0.019 \\
MS11$_{solar}$  & 3500 - 7429\AA & 2.54\AA & 6.5 Myr-15 Gyr & Z$=$0.02 (solar metallicity) \\
\hline\end{tabular}
   \begin{tablenotes}
      \small
      \item Notes: The \citet{BC03} model has a resolution of 3\AA\ from 3200--9500\AA, and lower resolution 
       in other wavelength ranges. G05 refers to the model published in \citet{G05}. 
      We used the Padova isochrone choice for the MILES and G05 models.
      We used the MILES based stellar population from MS11 at solar 
      metallicity only in this work to compare with \citet{Thomas13}.
    \end{tablenotes}
  \end{threeparttable}

\end{table}

We also note that the resolutions of different SPMs are different, but this does not have a significant impact on 
our results, since the SDSS spectra have lower or similar resolution as the template libraries. In addition,
 the velocity dispersions in host galaxies (a few hundred \kms) are bigger
 than the differences in resolution (a few tens \kms) of the stellar population models whose 
 resolution is determined by the stellar libraries used.
 When comparing the models with the observations, 
 broadening the stellar models to fit the spectral features of the observations erases the resolution differences.
 Therefore, the resolution  difference issue is not further discussed in this work.

\subsection{Spectral line features of stellar population models}

We first directly compare different stellar population templates from each SPM available with the same parameters (i.e., age, 
metallicity, and initial mass function), as the spectral line features are particularly important 
in the subtraction procedure to yield the type II AGN candidates.
In Figure ~\ref{fig:milesbc03full} we show the MILES and BC03 stellar population templates for
 representative ages of 63 Myr, 1 Gyr and 
12.5 Gyr at solar ($\rm [Fe/H]= 0.0$) and sub-solar metallicity ($\rm [Fe/H]=-0.4$). 
We see that for a given population, these two 
sets of models agree best at older ages (12.5 Gyr) with solar metallicity. At younger ages, i.e., 
63 Myr and 1 Gyr, inconsistencies are seen in both the Balmer lines and the overall shapes (i.e., the colors).
 At sub-solar metallicity, there are inconsistencies in the red, $\sim$ 6800--7400 \AA, especially for older 
 populations. Although we have normalised the 
models at 5500\AA, there are still some color discrepancies. 

\begin{figure}
\centering
\subfloat[]{\includegraphics[scale=0.39,angle=90]
    {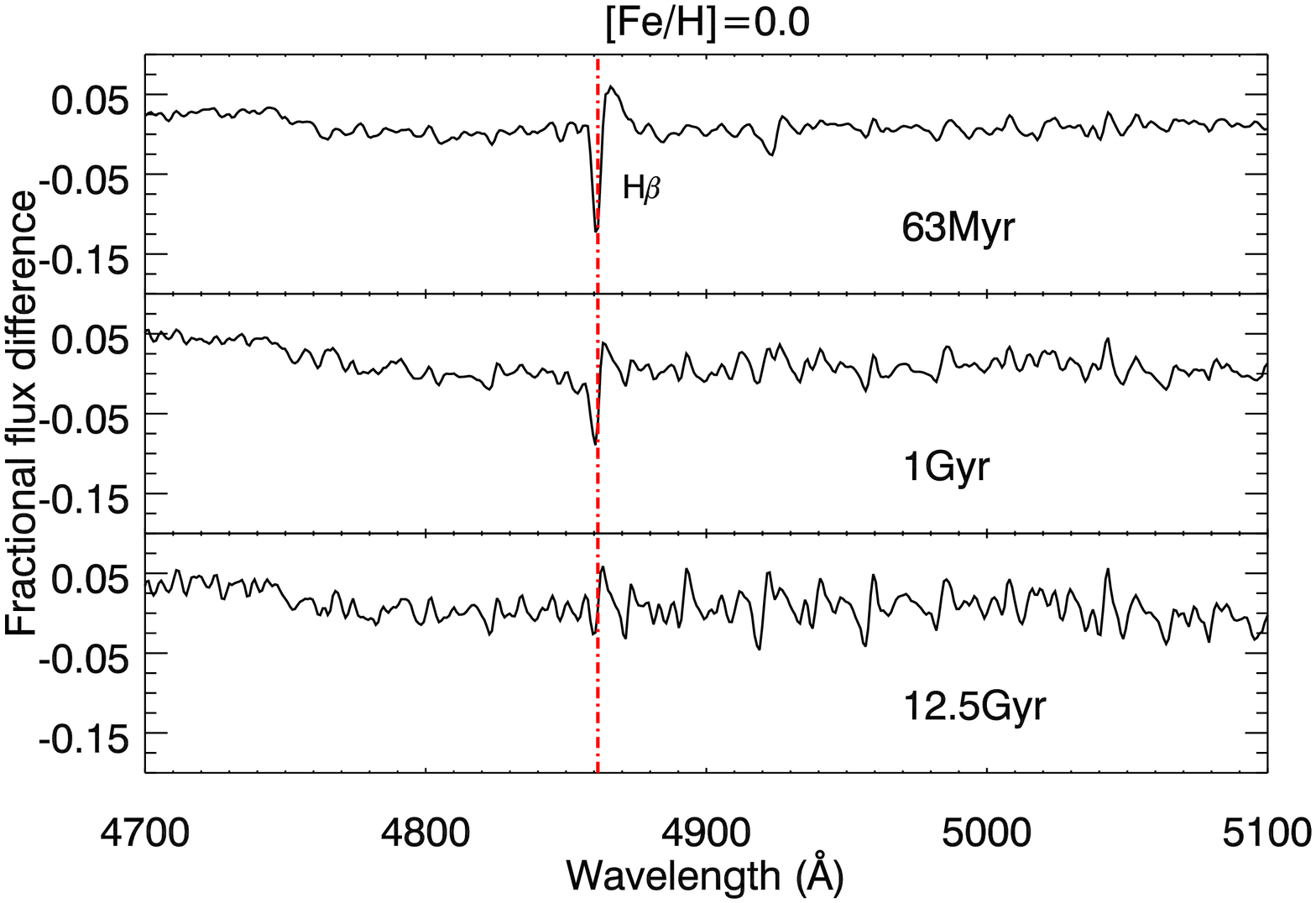}
\label{fig:milesbc03bluez00}} 
\hspace*{-4.0em}
\subfloat[]{\includegraphics[scale=0.39,angle=90]{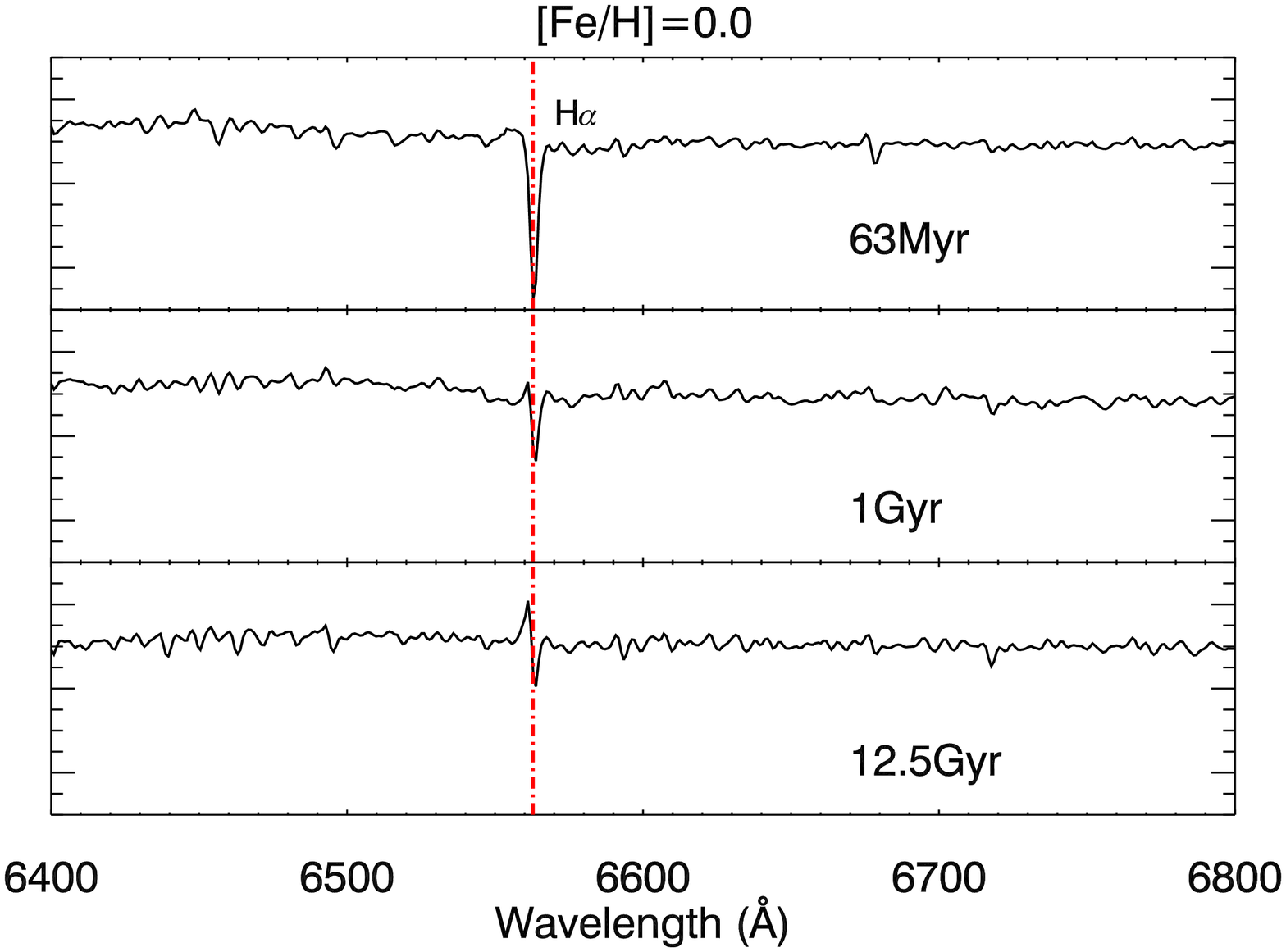}\label{fig:milesbc03redz00}}
\caption{Left: Model comparison zoomed in the blue  at solar metallicity ($\rm [Fe/H]=0.0 $), spectra were 
normalised at $\lambda \lambda$ 4830\AA. Right: Model comparison zoomed in the red at same metallicity, spectra were normalised at $\lambda \lambda$ 6520\AA. The fractional flux difference is derived by taking the difference between MILES models and
BC03, normalized to MILES models.}
\end{figure}

To better understand the similarities and differences of the models in the wavelength ranges of the four main type II AGN identification lines, we zoom in on the SSPs.
In Fig~\ref{fig:milesbc03bluez00}, we see 
that at solar metallicity, 
these two sets of models agree with each other 
especially at older ages (12.5 Gyr). However, models show differences in the depth of $\rm H{\alpha}$
(Fig~\ref{fig:milesbc03redz00}) at younger ages, with a 7\% of difference observed for 1 Gyr models.
In general, MILES templates show stronger absorption Balmer lines than BC03. 

\begin{figure}
\centering
\subfloat[]{\includegraphics[scale=0.39,angle=90]
    {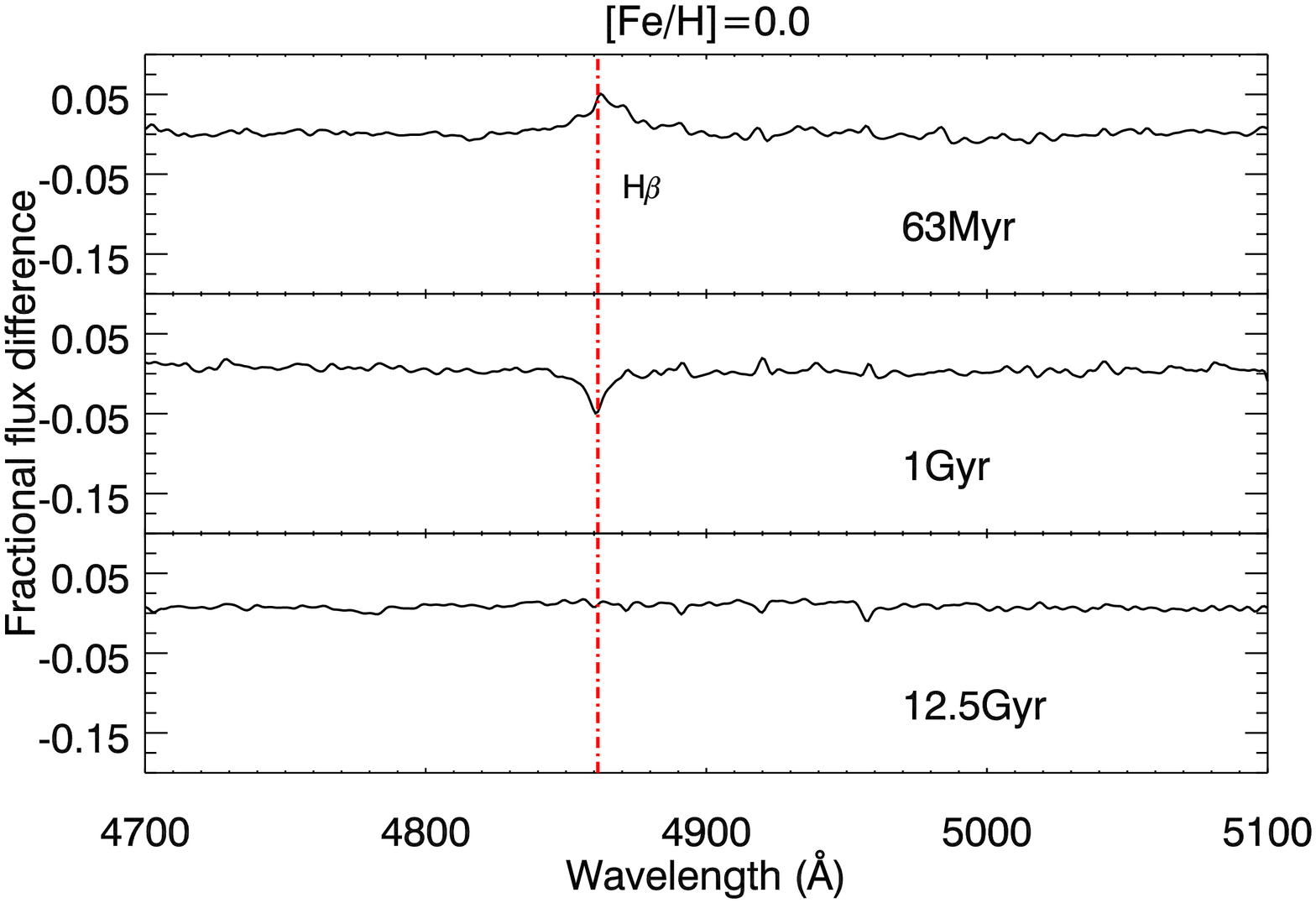}\label{fig:milesm11bluez00}} 
\hspace*{-3.6em}
\subfloat[]{\includegraphics[scale=0.39,angle=90]{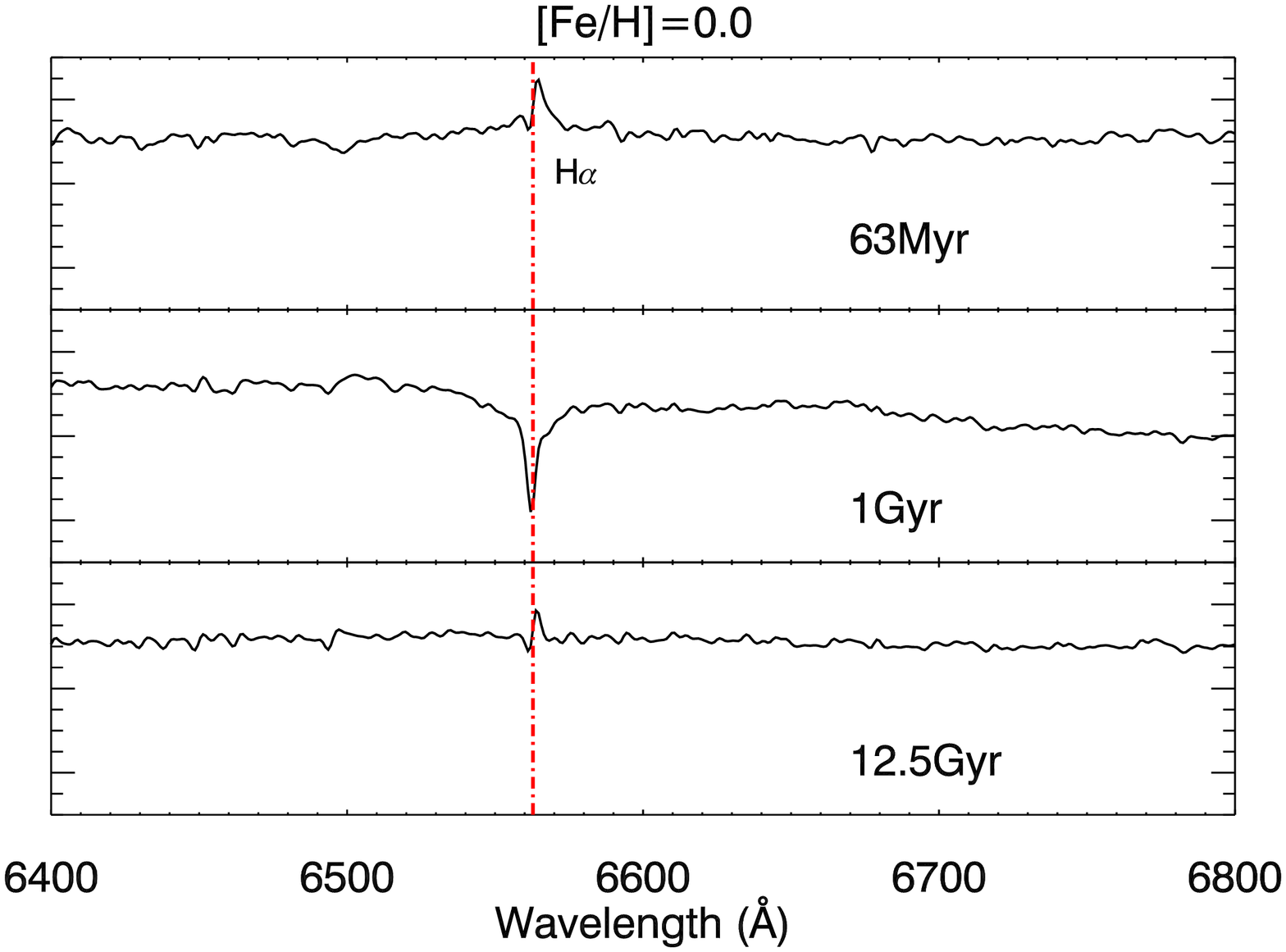}\label{fig:milesm11redz00}}
\caption{Solar metallicity ($\rm [Fe/H]=0.0 $) model comparison between MILES and MS11. Both 
models are smoothed to the same resolution as in BC03.  Left: Model comparison zoomed in for
 blue wavelength ranges. Spectra were normalised at $\lambda \lambda$ 4830\AA. Right: Model
 comparison zoomed in for red wavelength ranges. Spectra were normalised at $\lambda \lambda$ 
 6520\AA. The fractional flux difference is derived by taking the difference between MILES models and
MS11, normalized to MILES models.}
\end{figure}

We show the MILES-based MS11$_{solar}$ and compare them with 
MILES SSPs in Figs~\ref{fig:milesm11bluez00} and~\ref{fig:milesm11redz00}. 
 Although both MS11$_{solar}$ and MILES used the MILES 
 stellar library as their optical input, we still observe remarkable line differences and continuum variation, due to
the other inputs that comprise a SPM. 
The MS11 models tend to have slightly stronger Balmer absorption lines than MILES at the youngest and oldest ages, 
 but have weaker Balmer absorption lines than MILES at intermediate ages (1 Gyr).

\section{Is type II AGN identification template-dependent?}\label{sectest}

To check our method and the validity of our results, we compared our line fluxes with the values 
in the SDSS data release for the emission line galaxies identified by our MILES-based template subtraction. Note that
we use the MILES stellar population model while SDSS DR8 uses BC03; 
therefore some differences in emission line fluxes are expected. The comparison between the fluxes 
of diagnostic lines ($\rm {H}{\beta}$, \OIII\ 5007, $\rm {H}{\alpha}$ and \NII\ 6584) from 
our fits and those from SDSS shows a good linear correlation (see ZCF18). 
The $\rm {H}{\beta}$ line is the weakest in general, so the scatter is larger than in the other lines.

Although the line fluxes correlate well overall, AGN identification based on the line flux ratios 
{\OIII\ 5007/H}${\beta}$ and {\NII\ 6584/H}${\alpha}$ may have significant differences. Since
galaxies near the boundary are the most likely to shift categories (AGN to star-forming galaxies or vice versa),
we begin to examine differences in line ratio by selecting a subsample of objects within 0.02 dex of the \citet{K01} boundary
using MILES stellar population templates for host galaxy subtraction.
This subsample contains 145 emission line galaxies. 

 \subsection{Type II AGN identification variations}

Fig~\ref{fig:BPTth13miles} shows the BPT diagram for this boundary sample, with MILES-based line ratios marked as blue 
triangles, BC03-based line ratios \citep{Aihara11} shown as grey crosses, and the MILES-based MS11$_{solar}$ \citep{Thomas13}
result shown 
as magenta triangles. The line ratios using BC03 are systematically higher than
those using MILES. 
Evidently, more galaxies are classified as AGN when the background galaxy subtraction is
processed with BC03 templates, while the line ratios  from using MS11$_{solar}$ templates 
are systematically shifted into the composite region, i.e., between \citet{K01} and \citet{K03} boundaries. 
The systematic shift of line ratios from MS11$_{solar}$ templates are also seen in the histogram,
 especially in the line ratio of \OIII\ 5007/ H${\beta}$. BC03 templates result in higher  \OIII\ 5007/$\rm H{\beta}$
 ratios than MILES, while MS11$_{solar}$ templates result in lower {\OIII\ 5007/H}${\beta}$
 ratios than MILES. The line ratio of {\NII\ 6584/H}${\alpha}$ remains in the similar range.

\begin{figure}
\centering
\includegraphics[scale=0.7,angle=0]{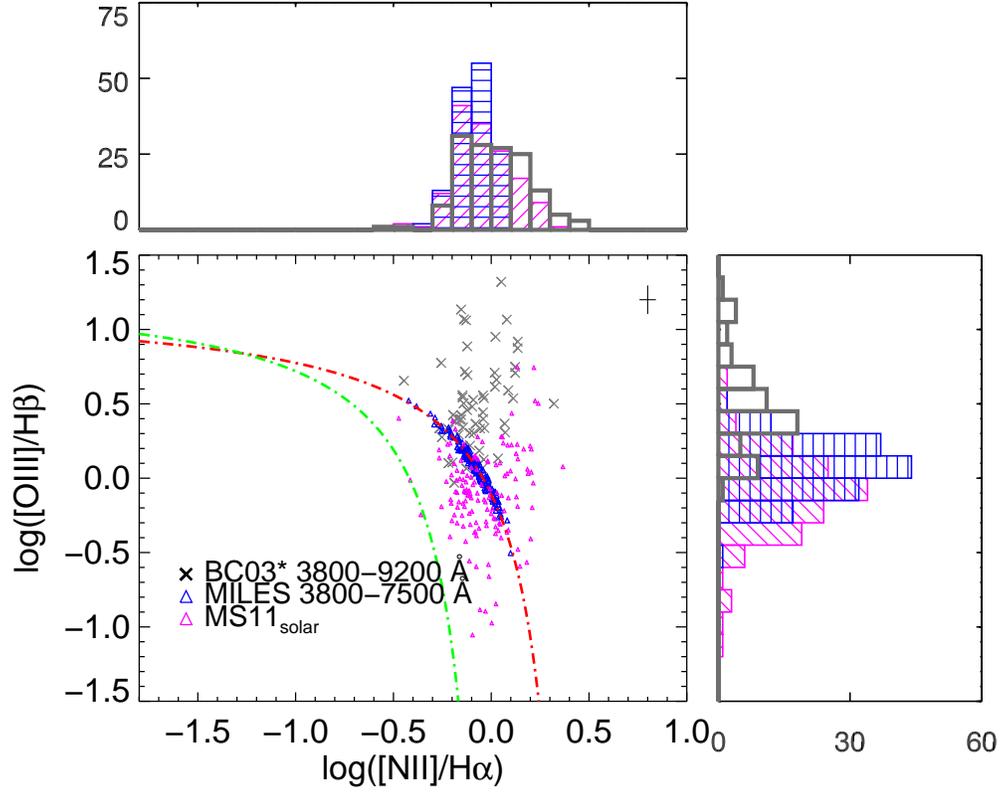}
\caption{
BC03-based line ratios (grey crosses) for 145 objects which MILES-based ratios (blue triangles) within 
0.02 dex of the \citet{K01} boundary. The MS11$_{solar}$  line ratios for the same galaxies (magenta triangles) 
systematically shift downwards. }
\label{fig:BPTth13miles}
\end{figure}

\begin{figure} 
\centering
\hspace*{-2.0em}
\subfloat[]{\includegraphics[scale=0.35,angle=0]
    {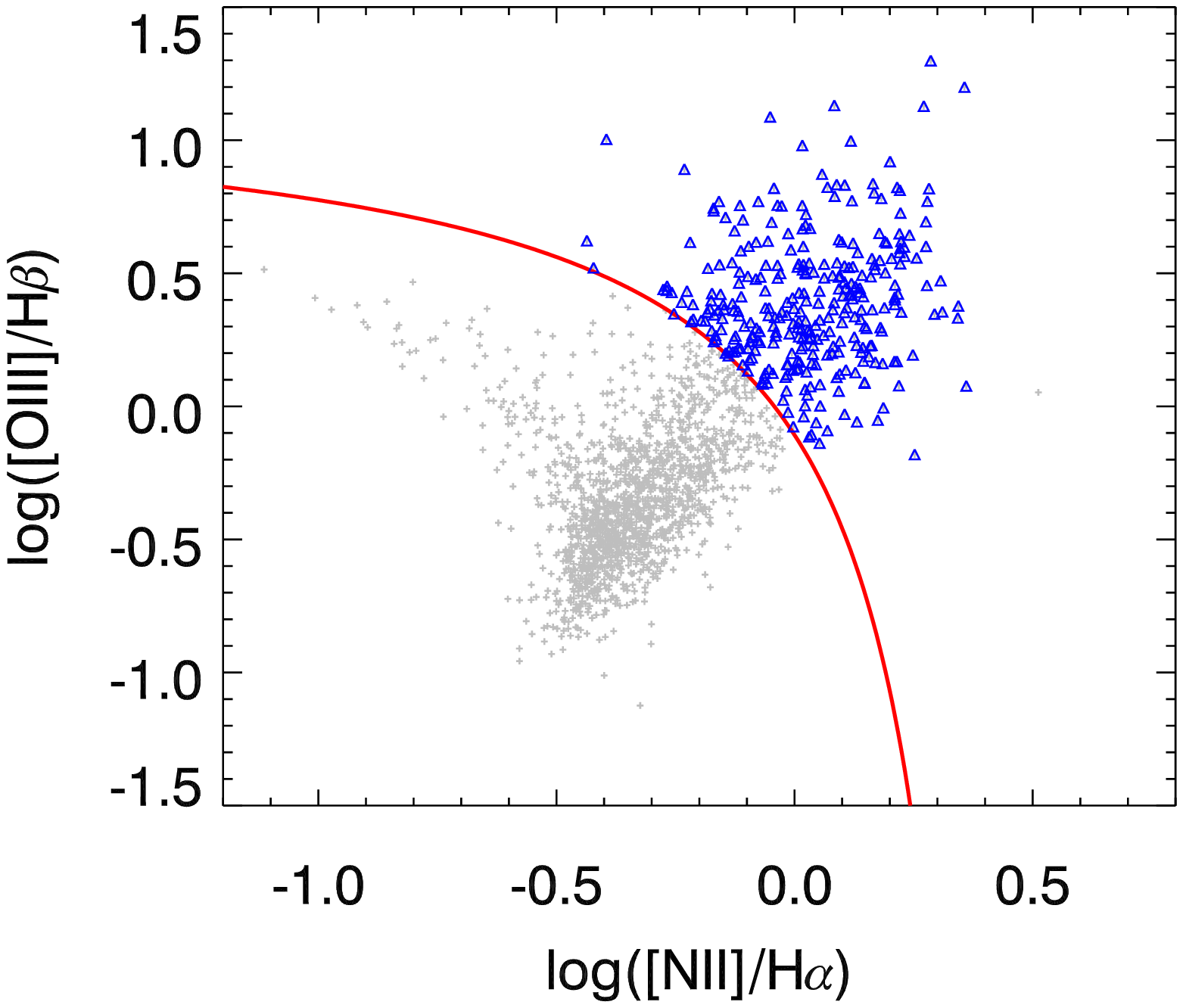}  }   
\hspace*{-3.72em}
\subfloat[]{\includegraphics[scale=0.35,angle=0]
    {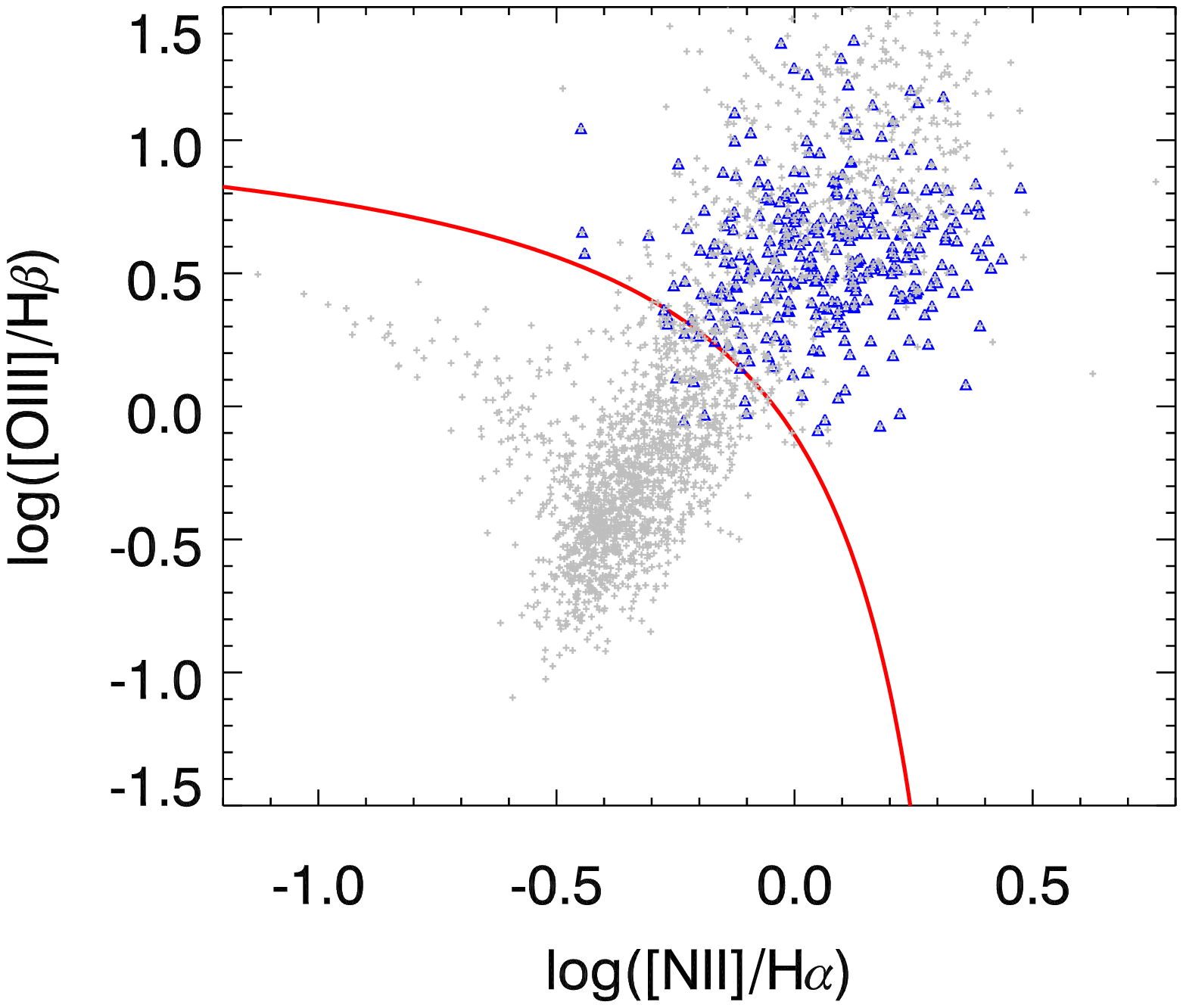}}
\hspace*{-3.72em}
\subfloat[]{\includegraphics[scale=0.35,angle=0]
    {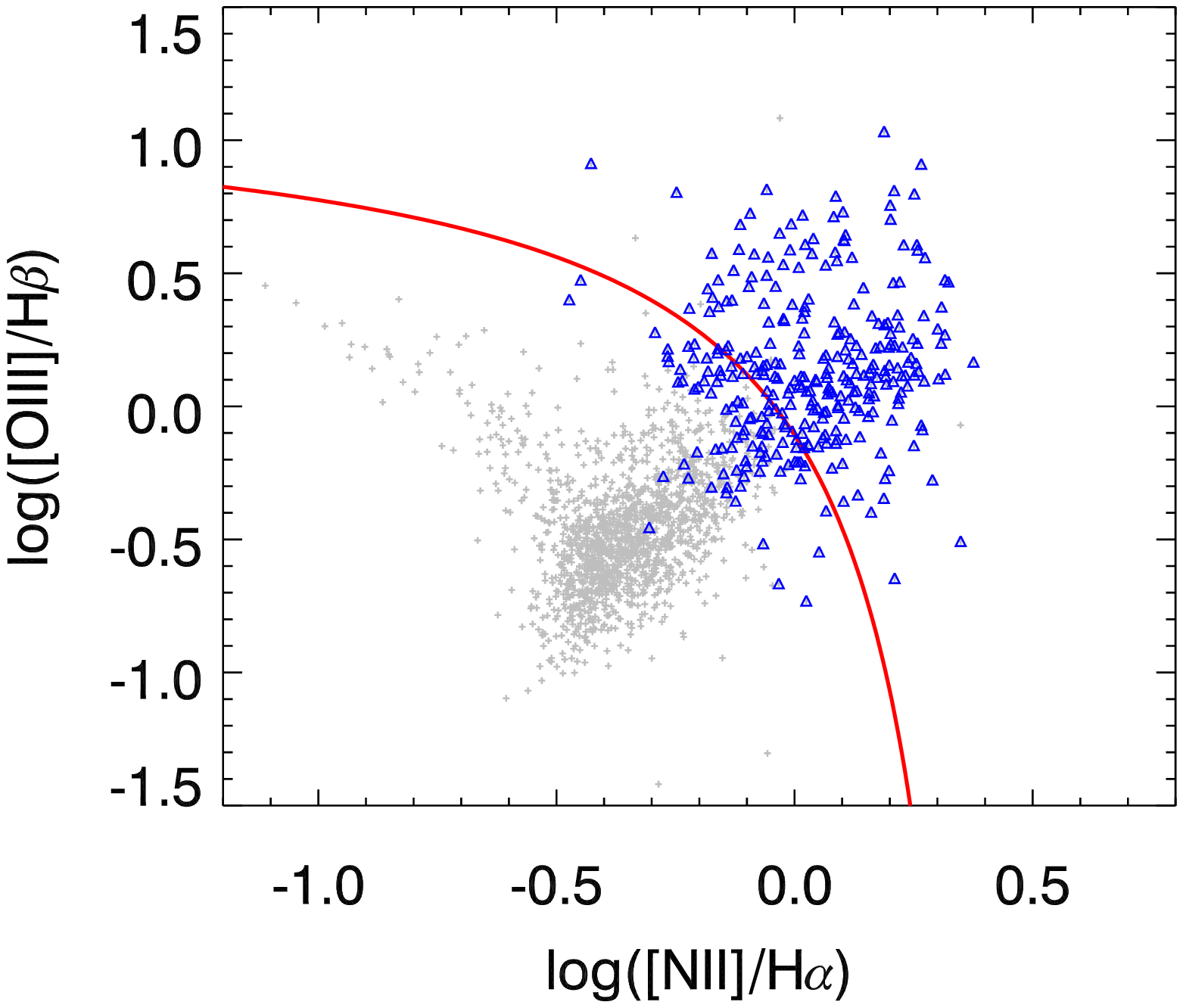}} 
\caption{Panel (a): Line ratios based on MILES template-subtraction to show the distribution of star-forming galaxies (grey crosses), and 
AGN (blue triangles). The data are restricted to have $\rm S/N \ge$3 for all four 
lines using the MILES template-subtraction.
Panel (b): Line ratios based on BC03 for the same galaxies in panel (a). 
Panel (c): Line ratios based on MS11$_{solar}$ for the same galaxies in panel (a).
}
\label{fig:bpt_normal_cmp_3temps}
\end{figure}

We then expanded our study to the full emission line sample defined by MILES template subtraction. 
The BPT diagram for these 3350 galaxies is shown in Fig~\ref{fig:bpt_normal_cmp_3temps}.
The narrow emission line galaxies are shown as grey crosses,
type II AGN (MILES based) are shown as blue triangles. 
In panels (b) and (c), we show the line ratios from fluxes when using BC03 and MS11$_{solar}$ 
templates, respectively, for the same galaxies. The symbols reflect AGN identification based on MILES template subtraction.
 We see that BC03-based line ratios
 tend to move toward the AGN region, except for a few outliers. MS11$_{solar}$ based line ratios
 move in the opposite direction, and some of the MILES-based AGN shift
into the composite region.

\begin{figure} [h]
\subfloat[]{\includegraphics[scale=0.39,angle=90]
    {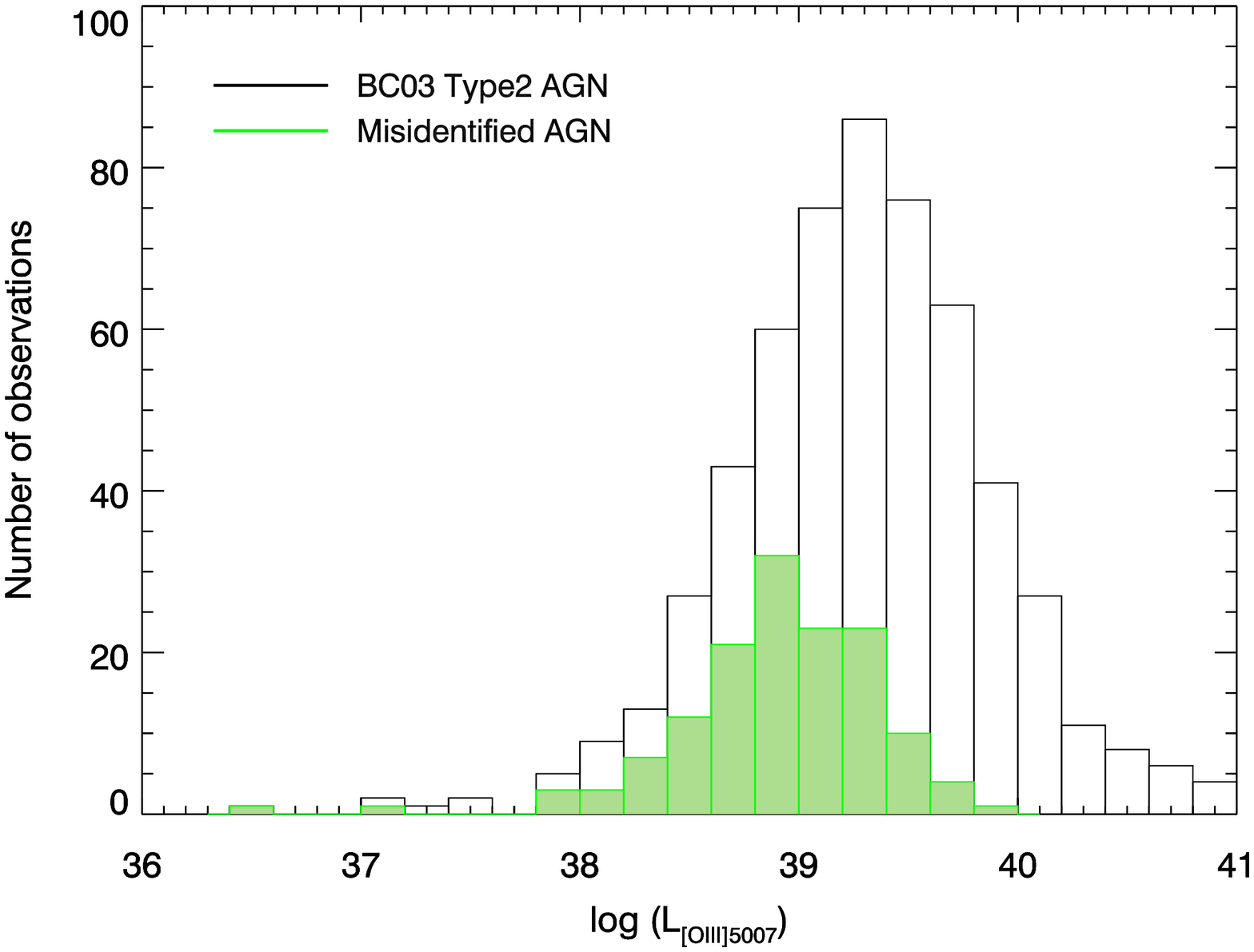}\label{fig:moveagn}  }           
\hspace*{-1.9em}
\subfloat[]{\includegraphics[scale=0.39,angle=90]{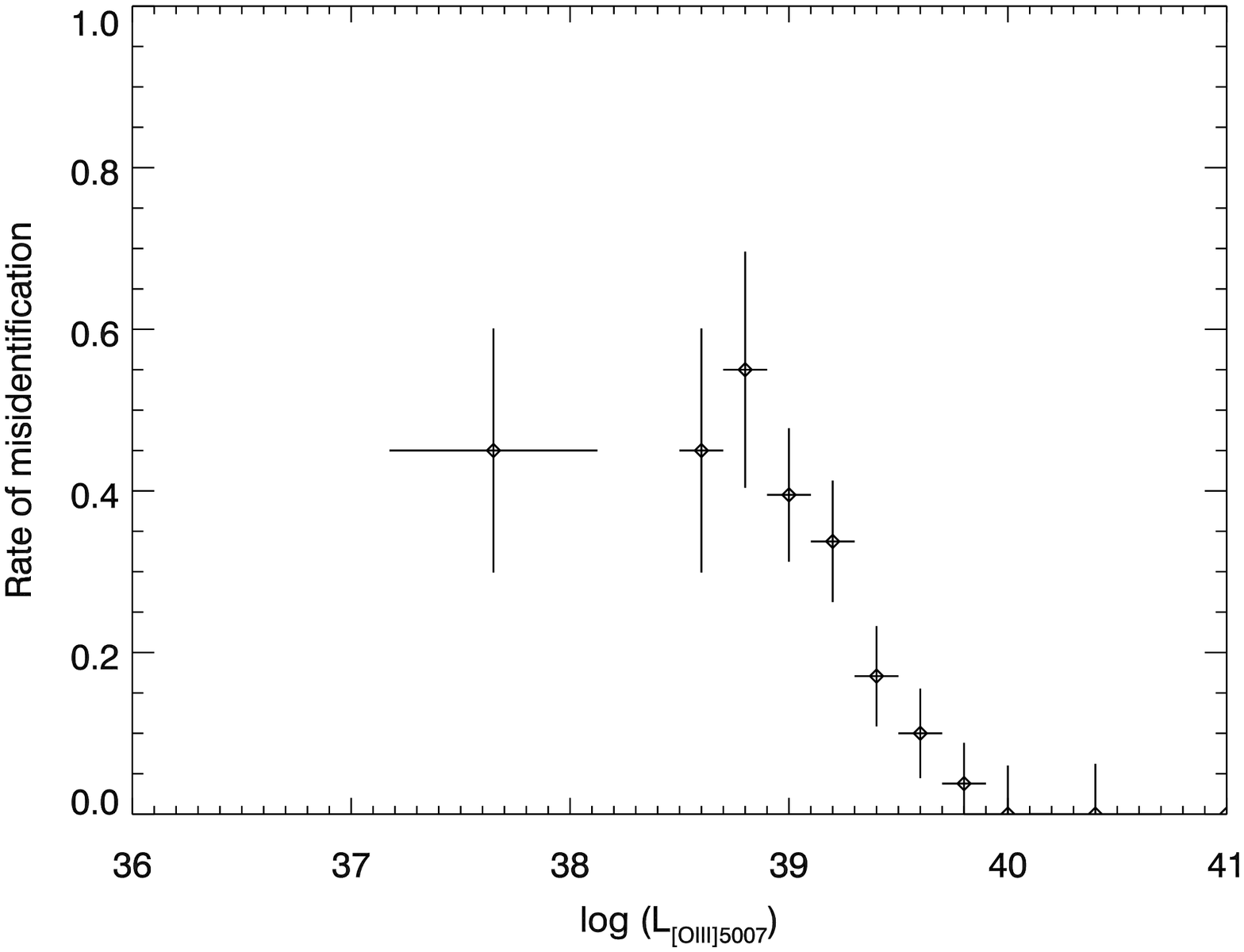}\label{fig:moveagnratio}}
\\
\subfloat[]{\includegraphics[scale=0.39,angle=90]
    {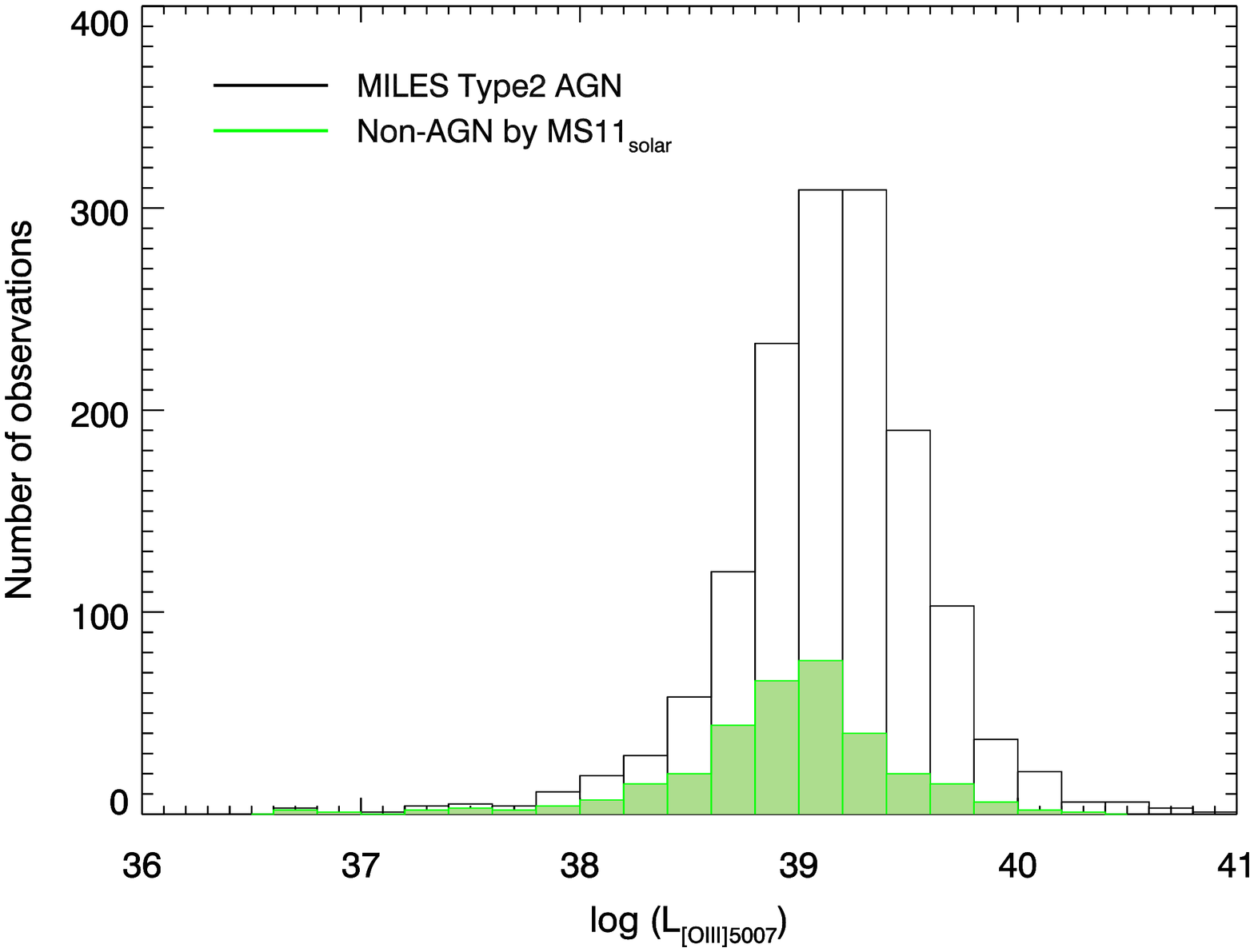}\label{fig:portsmoveagn}} 
\hspace*{-1.9em}
\subfloat[]{\includegraphics[scale=0.39,angle=90]{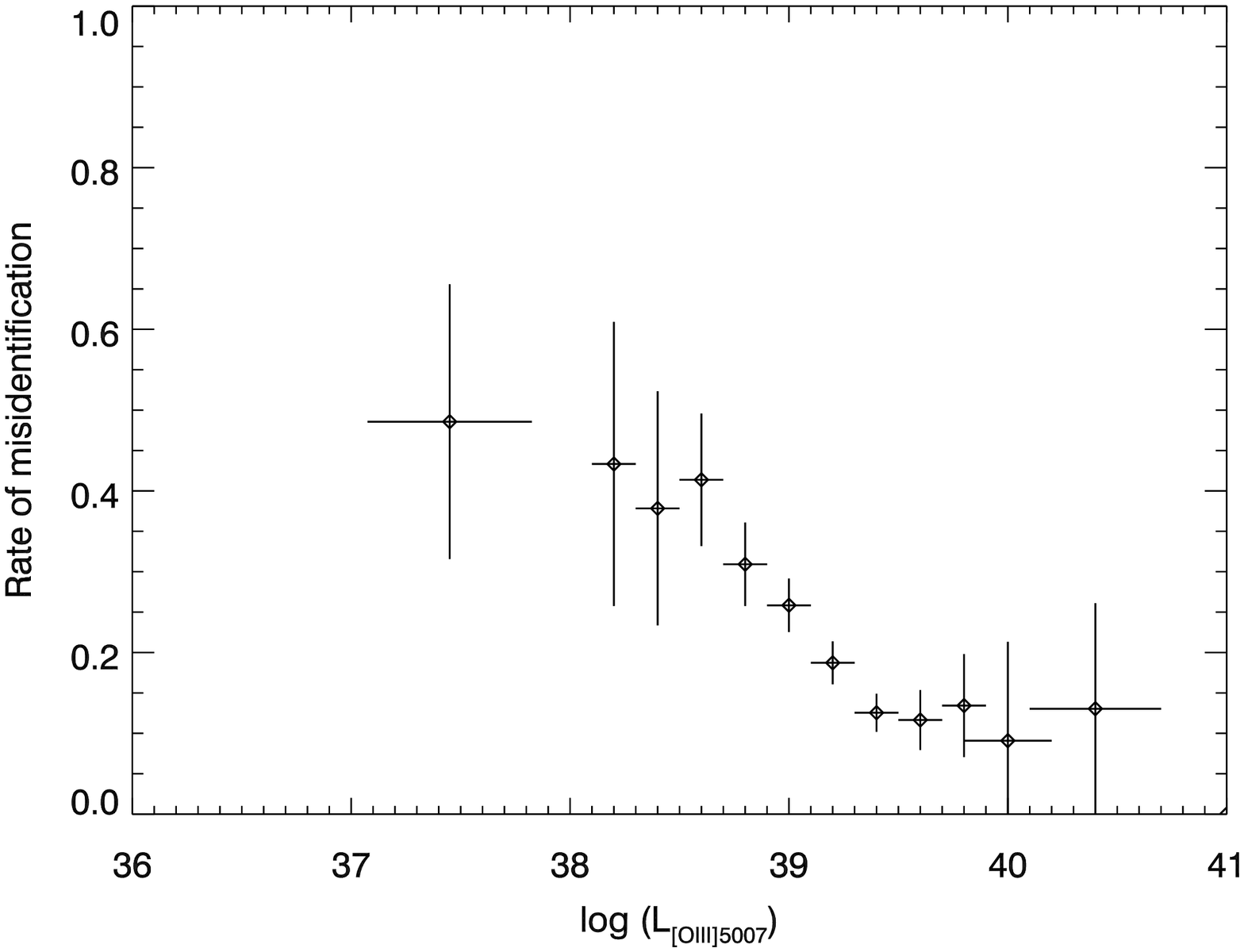}\label{fig:portsmoveratio}}

\caption{Panel (a): The clear histogram shows the \OIII5007 luminosity distribution of type II AGN identified by using
 BC03 templates. The objects not identified as AGN by using MILES templates are shown in green. Panel (b): 
 Rate of misidentified type II AGN using BC03 templates as a function of \OIII5007 luminosity. Panel (c): 
 The clear histogram shows the \OIII5007 luminosity distribution of type II AGN candidates identified by 
 using MILES templates. The objects not identified as AGN by the work of using MS11$_{solar}$ are shown 
 in green. Panel (d): Rate of misidentification of type II AGN by MS11$_{solar}$ as a function 
 of \OIII5007 luminosity.  }
\end{figure}

In order to learn which type II AGN are most sensitive to the choice of 
 stellar templates, 
we examine two classes of misidentification, taking the result from
  MILES template-subtraction as the standard. The first class is type II AGN identified by BC03 
 but not identified by MILES templates, i.e., ``false positives''; the second class is type II AGN identified by MILES 
 templates but not identified by MS11$_{solar}$, i.e., ``false negatives''. Since [\ion{O}{3}] 
 luminosity is an indicator of the AGN bolometric luminosity, we use it to determine whether the misidentification 
rate depends on AGN activity.
We show the \OIII\ luminosity distribution of BC03 type II AGN in Fig~\ref{fig:moveagn}. 
 The clear histogram shows the \OIII\ luminosity of 562 type II AGN identified 
 by using BC03 templates, selected from the MILES emission line galaxy sample; 
 the green histogram shows the false positives. We find that 25$\%$ of BC03 type II AGN 
are false positives\footnote{False negatives from BC03 and false positives from MS11$_{solar}$ are negligible (less than 2\%).}. 
 The fainter the object is, the more likely it is to be misidentified as an AGN. 
We show the ratio of misidentification of type II AGN candidates as a function of  \OIII\ luminosity 
in Fig~\ref{fig:moveagnratio}. The largest misidentification rate, $\sim 50 \%$, is found for galaxies with \OIII\ luminosity fainter than $\sim 10^{38}$ erg $s^{-1}$.

In Fig~\ref{fig:portsmoveagn}, type II AGN identified by MILES templates are shown in the clear histogram. 
Those false negatives when using line ratios from MS11$_{solar}$ template subtraction are shown 
in the green histogram. A total misidentification rate of $22\%$ is observed.
We plot the misidentification rate as a function of \OIII5007 luminosity
in Fig~\ref{fig:portsmoveratio}: objects with \OIII5007 luminosity fainter than $10^{38}$ erg $s^{-1}$ 
 can have a misidentification rate as large as $\sim 50\%$. 
 
 \begin{figure}
\centering
\includegraphics[scale=0.7,angle=90]{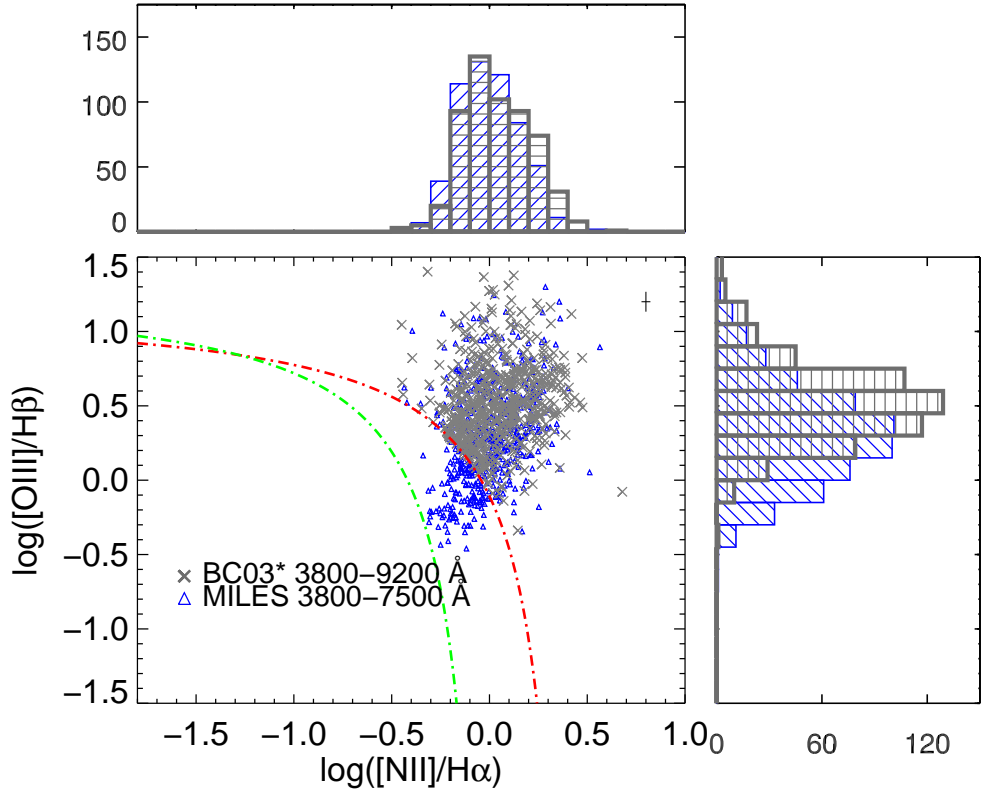}
\caption{
BC03-based line ratios (grey crosses) for 565 galaxies with S/N$\ge$3 above \citet{K01} boundary. MILES-based ratios (blue triangles) for the same galaxies systematically shift downwards. }
\label{fig:BPTbck01sn3milessn3}
\end{figure}

Furthermore, we limit the 2MRS-SDSS sample to have S/N$\ge$3 for all four AGN diagnostic lines 
for all three sets of templates
 to avoid effects of low signal to noise, which results in 2300 remaining galaxies.
 As shown in Fig~\ref{fig:BPTbck01sn3milessn3}, there is still a clear
 systematic shift between BC03 and MILES line ratios. We calculate the misidentification fraction with 
 the new sample and get a similar result:  21.5$\%$ 
 of type II AGN identified by using BC03 templates are not identified as AGN when using MILES templates.
 Similarly, 28.1\% of MILES type II AGN fall below the \cite{K01} boundary when using MS11$_{solar}$ derived fluxes.


\subsubsection{Misidentification of LINERs and Seyfert II}
Low-ionization narrow emission-line regions (LINERs) were first defined by \cite{Heckman80},
and are different from Seyfert II galaxies. Typical LINERs have spectra features dominated by lines arising from
low ionization states; their luminosities are similar to giant H{\sc ii} regions; and they are common.
AGN located above the \citet{K01} boundary in the 
 \OIII/$\rm H{\beta}$--\NII/$\rm H{\alpha}$ diagram can be either LINERs or Seyfert II. 
It is, therefore, pertinent to investigate if 
LINERs or Seyfert II dominates the misidentification rate.
In addition, $\rm H{\alpha}$ equivalent width (EW) $< 3$\AA\ is proposed to be an indicator of LINERs whose emissions are
from hot evolved stars \citep{Cidfernandes10}.
The detailed discussion is in Appendix~\ref{appmisliner}. In summary, no significant difference is found in the
misidentification rate between LINERs and Seyfert II. Since none of the misidentified Seyfert II or LINERs have 
$\rm H{\alpha}\ EW \leq 3\AA$, 
it is unlikely the misidentified LINERs in our sample are powered by hot evolved stars.

\subsubsection{Dependence on data quality}
Our study of the misidentified fraction as a function the overall quality of the spectrum, i.e., the continuum S/N ratio
is detailed in Appendix~\ref{appmissn}. In summary, no strong correlation between the misidentified fraction and 
data quality is found.

\subsection{Comparisons of Templates}

\begin{figure}
\plotone{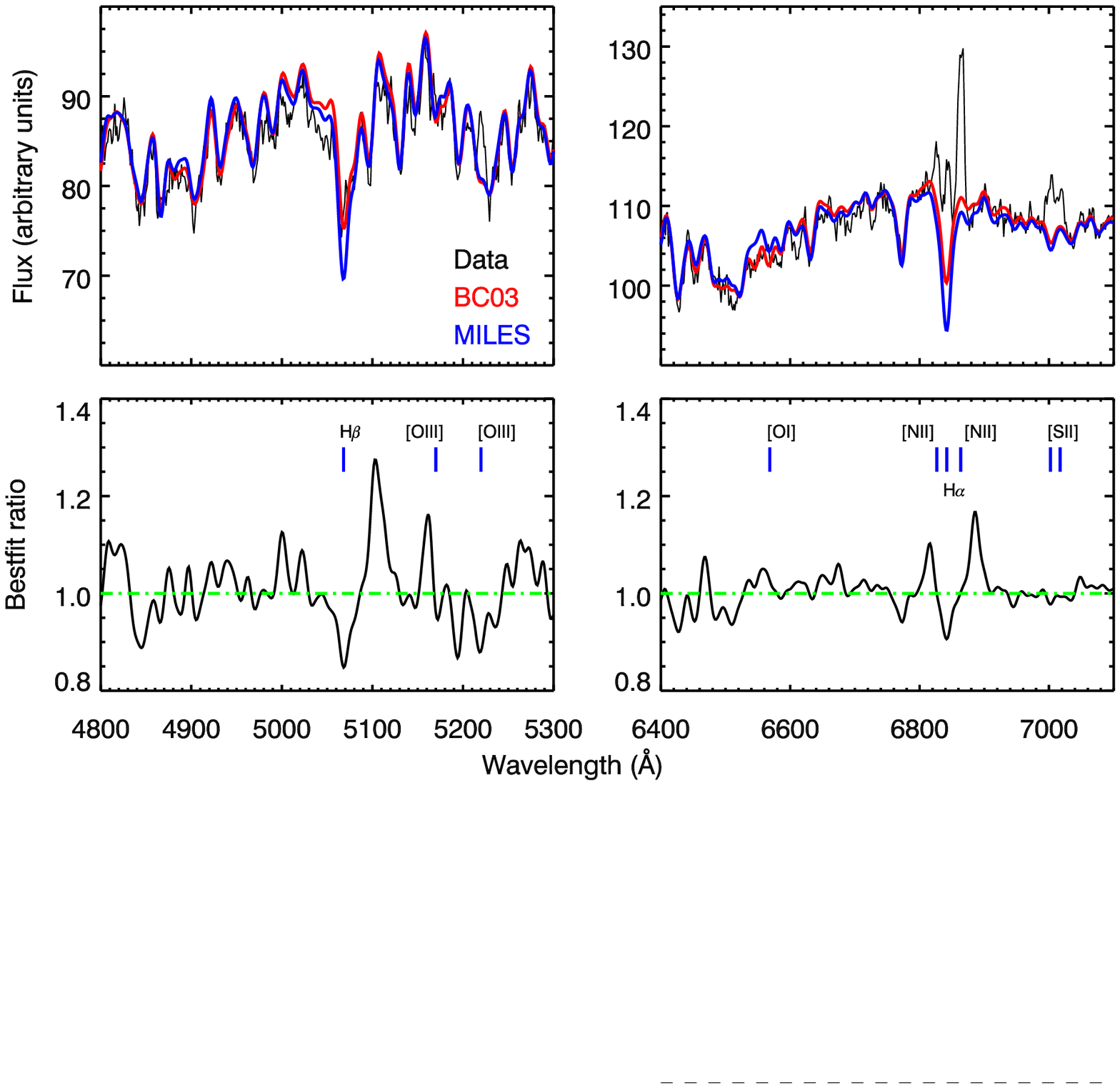}
\caption{An example showing comparison of the best fit models from the full-spectrum-fitting result using MILES and BC03 as templates. The `bestfit ratio' is the ratio between best fit derived by BC03 templates and the one derived by MILES templates. }
\label{fig:dr8w2models}
\end{figure}

\begin{figure}
\centering
\includegraphics[scale=0.5,angle=0]{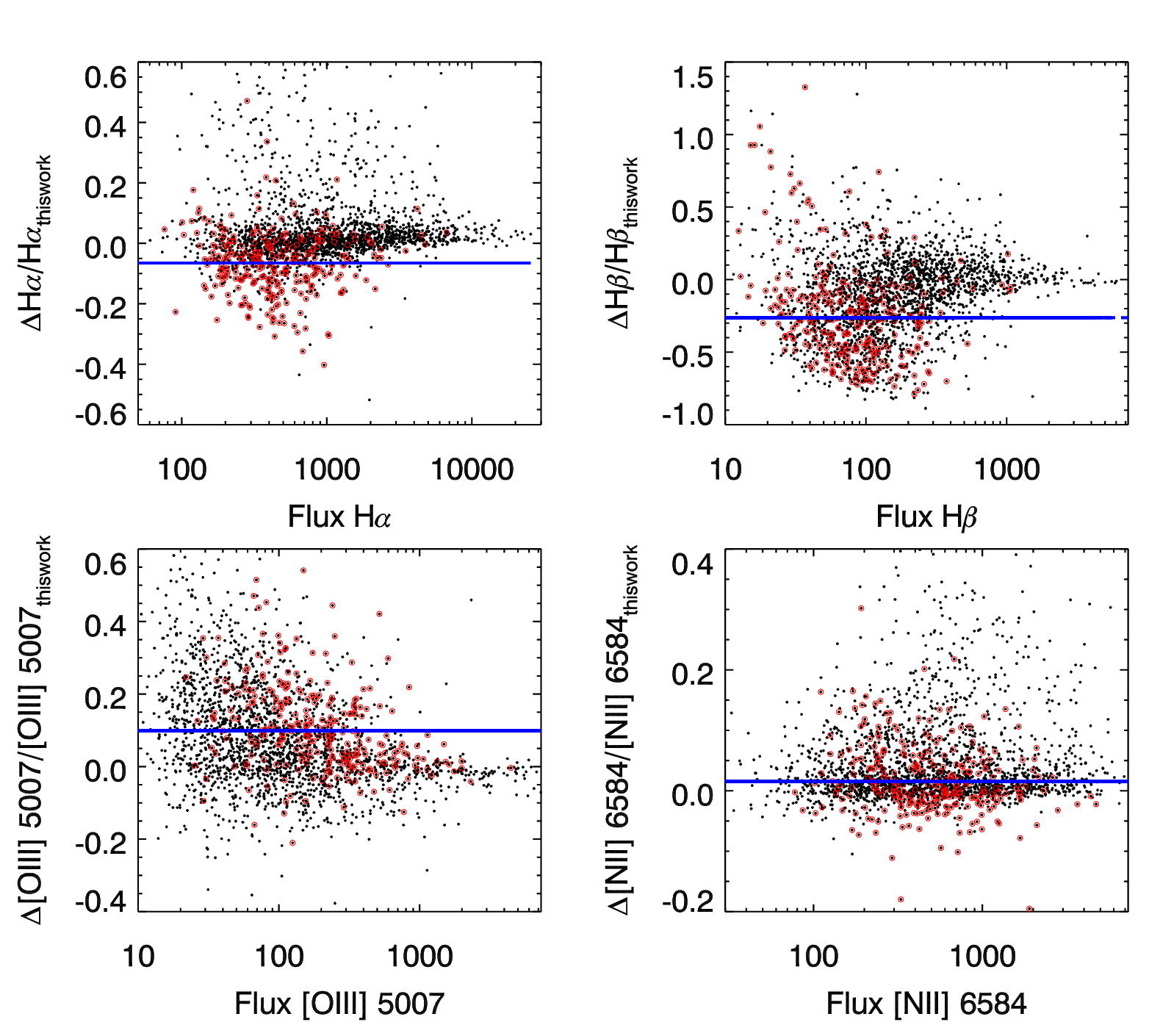}
\caption{The difference of line fluxes between BC03 and this work (using MILES). 
This work, BC03, and MS11$_{solar}$ of the 2300 objects with S/N$\ge$3 of all four 
lines from both templates are shown in black crosses. Type II AGN (i.e., galaxies with line ratios above \citet{K01} boundary 
in \OIII/$\rm H{\beta}$--\NII/$\rm H{\alpha}$ diagram) based on MILES template subtraction are
shown as red circles. Note the line ratio scatters are large. We mark mean values of the line ratios of the type II AGN 
as blue lines in each panel.
}
\label{fig:fluxes_bc03_milesk01_diff}
\end{figure}

Fig~\ref{fig:dr8w2models} illustrates the difference in the best fit  using the MILES and BC03 stellar population
models as templates, zooming in on the best-fit ratios
around the line regions we use to identify type II AGN candidates. 
The best fit using the MILES model predicts stronger Balmer lines (i.e., H${\alpha}$ and H${\beta}$ lines) than 
that using BC03 templates. 
We show the flux variations between the results from BC03 and MILES template subtraction in Fig~\ref{fig:fluxes_bc03_milesk01_diff}.  
The type II AGN identified by MILES template subtraction are highlighted in red circles.
The mean flux of H${\beta}$ from BC03 type II AGN is $\sim 26\%$ fainter than that from MILES templates, and the H${\alpha}$ line fit from BC03 is $\sim6\%$ fainter.
Most of the type II AGN have stronger \OIII \ fluxes from BC03 than from MILES templates. 
The fluxes of \NII\ from BC03 and MILES templates are consistent with each other. 
The inaccurate absorption lines of BC03 model is also described in detail by other groups \citep[e.g.,][]{G05, Koleva08}.

\begin{figure}[h]
\centering
\includegraphics[scale=0.5,angle=0]{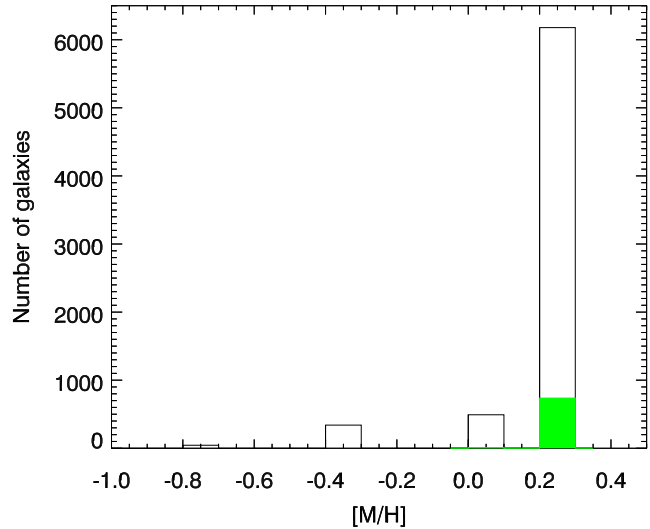}
\caption{ The metallicity distribution of the major population components of SDSS DR8 galaxies from
\textbf{the} full-spectrum-fitting result. The green histogram shows the ones that contain only one simple stellar population. }
\label{fig:ssp_major}
\end{figure}

The differences between MILES and MS11$_{solar}$ are more surprising since they use the same input
stellar library. This difference could be due to the treatment of the TP-AGB stars affecting SSPs between
0.2 to 2 Gyr. However,
we examined the stellar population components of the sample galaxies, and found only $8\%$ of them contain
younger (age $\le$ 2 Gyr) populations. In each of these galaxies, young stellar populations contribute less 
than 56\% of the optical light.
This means the systematic shift of line ratios when using MS11$_{solar}$
is not dominated by the special treatment of TP-AGB stars. 

Furthermore, we note that we used all the available metalicities in 
the MILES model, and our best fits indicate that most ($\sim70\%$) of our galaxies favour  
metal-rich models, as shown in Fig~\ref{fig:ssp_major}. 
The green subsample high-lights galaxies containing only one stellar population. 
While it is true that there is a degeneracy between age and metallicity when fitting only colors, as
noted by \citet{Thomas13}, the degeneracy is lifted when simultaneously fitting both the absorption lines and the continuum
 \citep[e.g.,][]{Reichardt01}.
Therefore, the choice of metallicity range of the templates also plays a role in type II AGN identifications.
A detailed comparison between the stellar population models by \cite{Maraston11} and 
\cite{Vazdekis10} is beyond the scope of this work, and will not be addressed here.

\subsection{Young stellar populations and wavelength range}

Young stellar populations ($\le$63 Myr) are absent in all of the templates families discussed above. 
However, as we show in the Appendix~\ref{appyoungssp}, the 
lack of young stellar populations does not significantly affect AGN identifications.
We use G05 to explore
young populations since empirical young SPMs are not available, and discuss 
the consistency between the G05 model and the MILES model at 63 Myr. An expansion of MILES 
model with G05 young population models is presented and used for AGN identification around
the \citet{K01} border. We also examined differences due to wavelength ranges used in host galaxy subtraction
in Appendix~\ref{appwavelength}. 
We find that young stellar populations (i.e. age $\le$ 63Myr) and the wavelength ranges used in the fit
make little to no difference in AGN identification rates. 

\section{Conclusions}\label{secconclusion}

We have examined the differences in optical identification of type II AGN 
in nearby ($z\le0.09$) galaxies with SDSS DR8 spectra, 
resulting from 
host galaxy subtraction using MILES, BC03, and MS11$_{solar}$ 
as stellar templates. We found  that 
type II AGN identification is sensitive to the stellar template. 
Comparing the results of using the BC03 and MILES SPMs to subtract absorption lines in 
SDSS DR8 data, we determined that one quarter of the sample is misidentified as type II AGN by 
BC03 relative to MILES. 
Results using the MS11$_{solar}$ templates show fewer galaxies identified as AGN relative to MILES.
We also find a $22\%$ disagreement overall the work of \citet{Thomas13}, 
which used MS11$_{solar}$ for their host galaxy continuum and absorption subtraction. 
We traced the problem to the incomplete range of metallicities of the SPMs used in template fitting.
The misidentification of both using BC03 (e.g., the work of MPA-JHU, SDSS DR8) and 
using MS11$_{solar}$ (e.g., the work of \citet{Thomas13}) is greatest for objects with low \OIII\ luminosities, 
and is up to $50\%$ for \OIII5007 luminosity fainter than $10^{38}$ erg $s^{-1}$.
The stellar population models used for the subtraction of the host galaxy contribution should be taken into account when 
using the emission line fluxes, or the AGN fractions, from a catalog especially if the results from different catalogs are compared.


\acknowledgments
\section*{acknowledgments}
We thank YuXiao Dai who helped us in checking the data and some useful discussions. We thank Jong-Hak Woo
for the discussion about AGN luminosities and AGN identifications. 
We thank the referee for his/her valuable comments and suggestions.
The research of GRF was supported by National Science Foundation grants NSF-PHY-1212538 and NSF-AST-1517319, and by the James Simons Foundation.

Funding for SDSS-III has been provided by the Alfred P. Sloan Foundation, the Participating Institutions, the National Science Foundation, and the U.S. Department of Energy Office of Science. The SDSS-III web site is http://www.sdss3.org/.

SDSS-III is managed by the Astrophysical Research Consortium for the Participating Institutions of the SDSS-III Collaboration including the University of Arizona, the Brazilian Participation Group, Brookhaven National Laboratory, Carnegie Mellon University, University of Florida, the French Participation Group, the German Participation Group, Harvard University, the Instituto de Astrofisica de Canarias, the Michigan State/Notre Dame/JINA Participation Group, Johns Hopkins University, Lawrence Berkeley National Laboratory, Max Planck Institute for Astrophysics, Max Planck Institute for Extraterrestrial Physics, New Mexico State University, New York University, Ohio State University, Pennsylvania State University, University of Portsmouth, Princeton University, the Spanish Participation Group, University of Tokyo, University of Utah, Vanderbilt University, University of Virginia, University of Washington, and Yale University.

\newpage
\appendix

\section{Misidentification of LINERs and Seyfert II}\label{appmisliner}

\begin{figure}
\centering
\includegraphics[scale=0.5,angle=90]{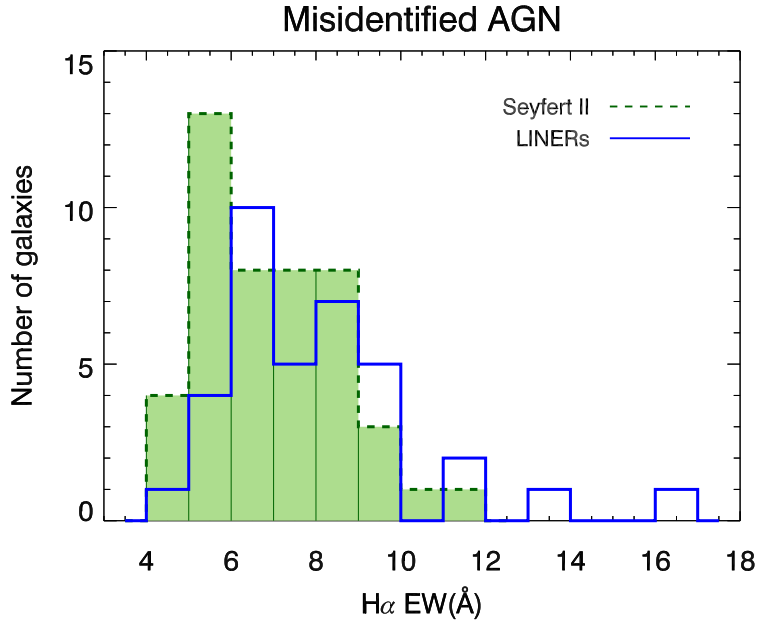}
\caption{$\rm H{\alpha}$ EW distribution of misidentified Seyfert II and LINERs when using BC03 templates.  
The clear histogram shows distribution of the misidentified LINERs and the filled green histogram shows the $\rm H{\alpha}$ EW
distribution of the misidentified Seyfert II.
}
\label{fig:bclinersy2_haew}
\end{figure}

Whether low-ionization nuclear emission-line regions (LINERs) are AGN is a topic
that has been debated in literature. \citet{Ho96, Ho08} and \citet{Masegosa11}, for example, argued that a significant fraction of LINERs are 
low-luminosity AGN. Other studies have suggested that LINERs are, instead, shock heated gas \citep{Dopita95},
starburst activity \citep{Terlevich85, AlonsoHerrero00}, or post-AGB stars \citep{Singh13}. 
\citet{Cidfernandes10} suggest that $\rm H{\alpha}$ EW can
differentiate between the different ionization mechanisms which lead to the overlap in the LINER region of
traditional diagnostic diagrams. According to the bimodal distribution of $\rm H{\alpha}$ EW, \citet{Cidfernandes11} suggest 
that LINERs with $\rm H{\alpha}$ EW $>$ 3 \AA\ are likely to be true AGN, while those with  
$\rm H{\alpha}$ EW $<$ 3 \AA\ have emissions from hot evolved stars.

In order to discern the nature of the LINERs in our sample using the  $\rm H{\alpha}$ EW, we first separate
the AGN identified from MILES based line ratios into LINERs and Seyfert II using the \citet{Kewley06} criteria.
Fig~\ref{fig:bclinersy2_haew} shows the misidentified LINERs and Seyfert II when using BC03 templates
as a function of $\rm H{\alpha}$ EW. None of the galaxies are found to have $\rm H{\alpha}$ EW $<$ 3 \AA. 
Therefore, there are no LINERs powered by hot evolved stars, as defined by \citet{Cidfernandes10}, in our misidentified sample.
The misidentification instead mainly comes from line ratio variations due to template subtraction. 
Furthermore, Fig~\ref{fig:bclinersy2_haew} shows that no clear separation of $\rm H{\alpha}$ EW distribution is observed
 between the misidentified LINERs and Seyfert II galaxies. 

\begin{figure} 
\centering
\hspace*{-2.0em}
\subfloat[]{\includegraphics[scale=0.35,angle=0]
    {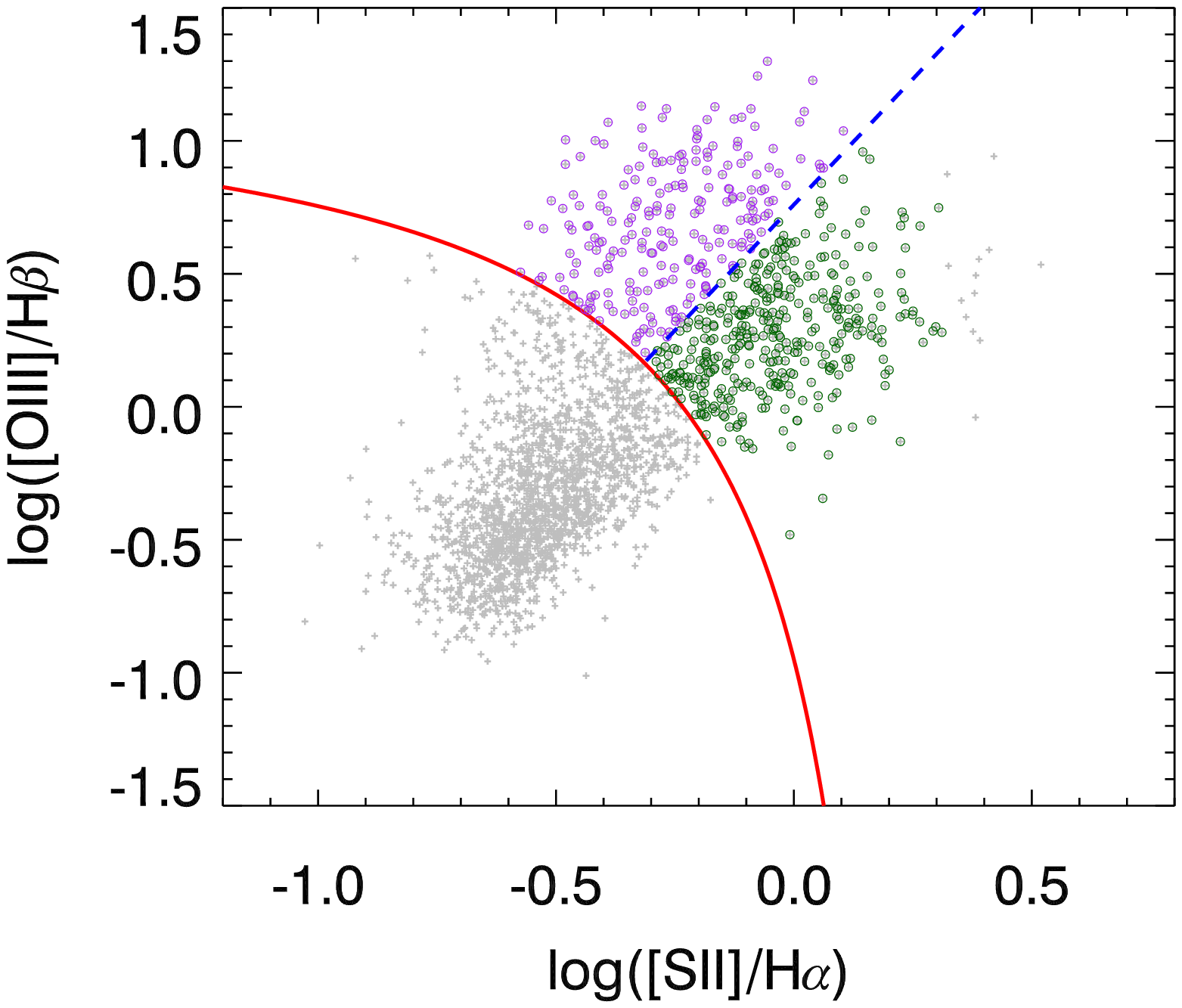}\label{fig:sub1}  }   
\hspace*{-3.72em}
\subfloat[]{\includegraphics[scale=0.35,angle=0]
    {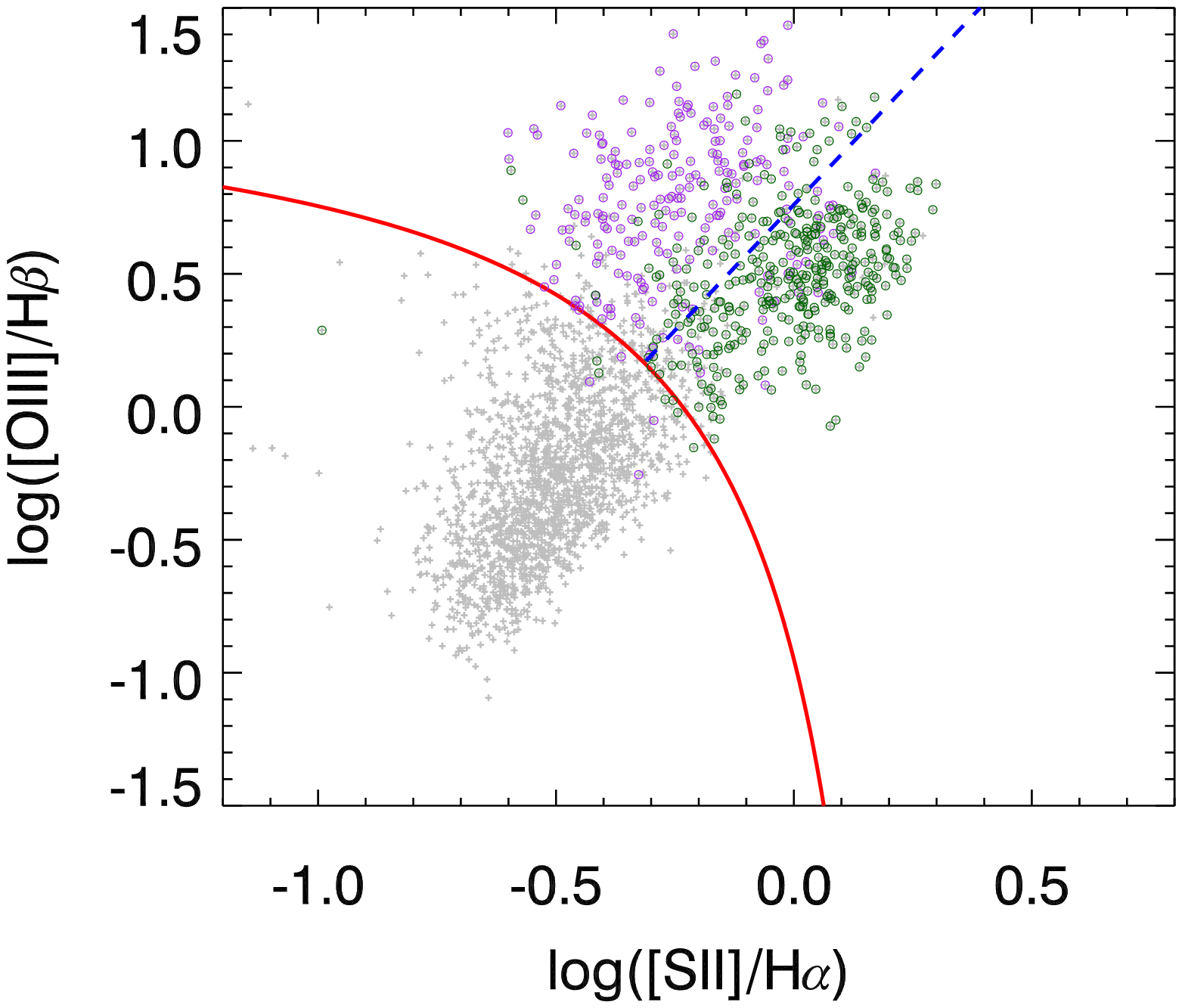}\label{fig:sub2}}
\hspace*{-3.72em}
\subfloat[]{\includegraphics[scale=0.35,angle=0]
    {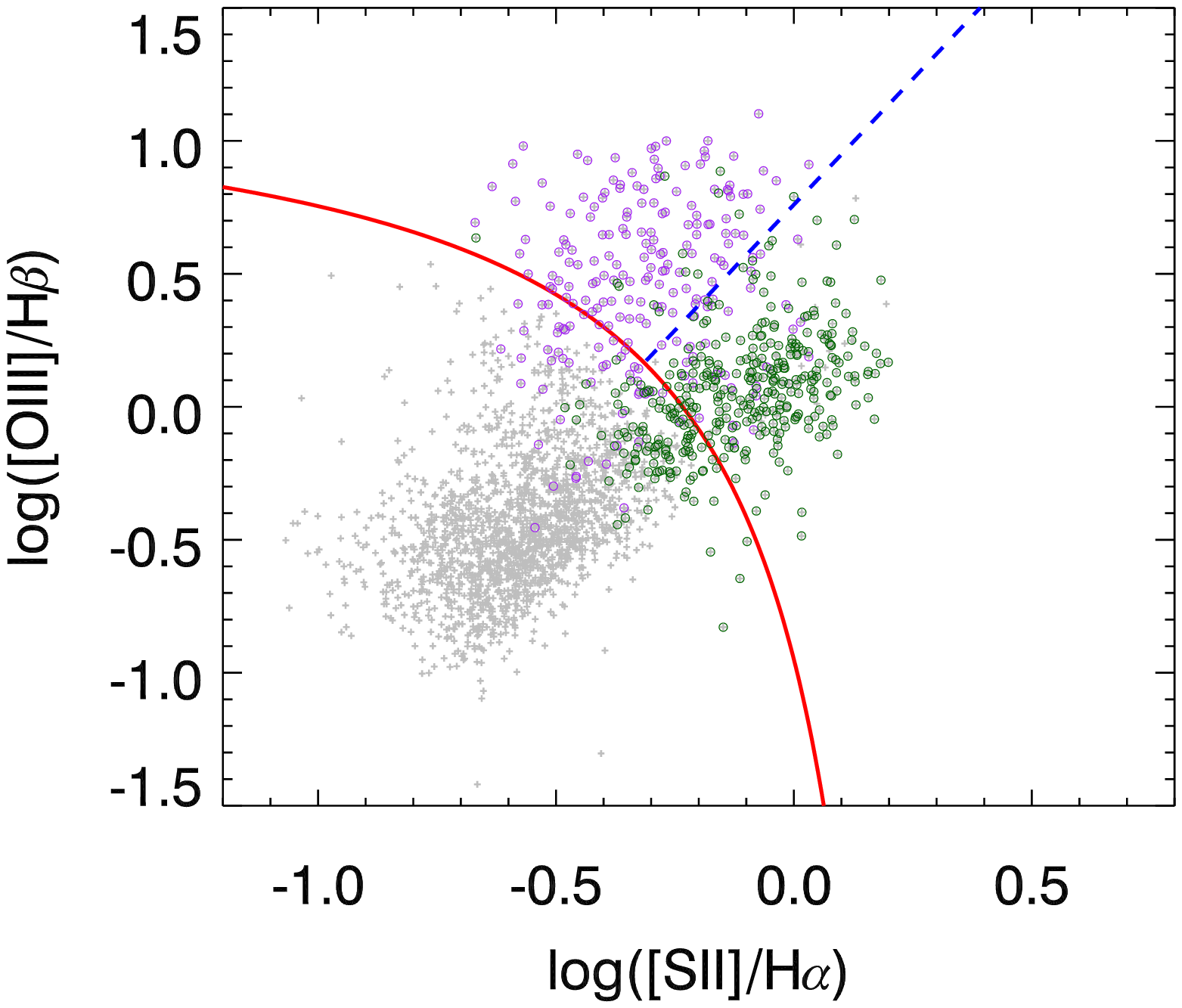}\label{fig:sub3}} 
\caption{Panel (a): Line ratios based on MILES template-subtraction to show the distribution of star-forming galaxies, Seyferts, 
and LINERs. $\rm S/N \ge 3$ were set up for all four lines from all three templates. 
The \citet{Kewley06} boundary (the blue dashed line) is adopted to distinguish Seyferts and LINERs, where Seyferts are 
 shown as purple circles, and LINERs are shown as green circles.  Panel (b): Line ratios based on BC03 
 for the same galaxies in panel (a). Panel (c): Line ratios based on MS11$_{solar}$ 
  for the same galaxies in panel (a).
The colors for panels (b) and (c) are based on MILES classifications.}
\label{fig:bpt_s2_liner_sy2}
\end{figure}

The line ratios based on MILES template-subtraction is shown in Fig~\ref{fig:bpt_s2_liner_sy2} panel (a).
Galaxies are limited to the sample with $\rm S/N \ge$ 3 for all four lines in all three templates.
Seyfert II are shown in purple and LINERs are shown in dark green. 
We also show the line ratios based on BC03 in panel (b) and line ratios based on 
MS11$_{solar}$ in panel (c) for the same galaxies. As clearly shown in the plots, line ratios derived from the different
template-subtractions show systematic shifts. BC03 based line ratios shift towards the AGN/Seyfert II regions;
while the MS11$_{solar}$ based line ratios show the opposite trend.

Of the Seyfert II identified using BC03,  20.7\% fall below the \cite{K01} using MILES-based line ratios. 
Similarly, the false positive rate of the BC03 LINERs is 14.6\%.
Our investigation of MS11$_{solar}$ based line ratios shows a 26.0\%  false negative rate 
for Seyfert II and a 27.7\% false negative rate for LINERs.
Roughly half of the misidentified galaxies are (were) Seyfert II and half are (were) LINERs.

\section{Dependence on data quality}\label{appmissn} 
We examine the misidentified AGN fraction against the overall data quality of the spectrum, i.e., the 
continuum S/N ratio. Since S/N varies at different wavelengths, we choose the continuum S/N near the $\rm {H}{\beta}$ 
line (hereafter S/N($\rm {H}{\beta}$)) for this purpose as $\rm {H}{\beta}$ is the weakest of the four type II AGN 
identification lines. 
Figure~\ref{fig:bcsnhba} shows the distribution of S/N($\rm {H}{\beta}$) for type II AGN. The green histogram highlights the 
type II AGN identified when using BC03 templates, but not as type II AGN, when using MILES templates. The misidentification 
rate shows a weak dependence of S/N($\rm {H}{\beta}$) as seen in Fig~\ref{fig:bcsnhbb}. 
Exploration of the dependence between \OIII5007 luminosity and S/N($\rm {H}{\beta}$) is shown
in Fig~\ref{fig:bcsnhbc}. Misidentified type II AGN are shown as filled red triangles. No strong correlation is found.

\begin{figure} 
\subfloat[]{\includegraphics[scale=0.39,angle=90]
    {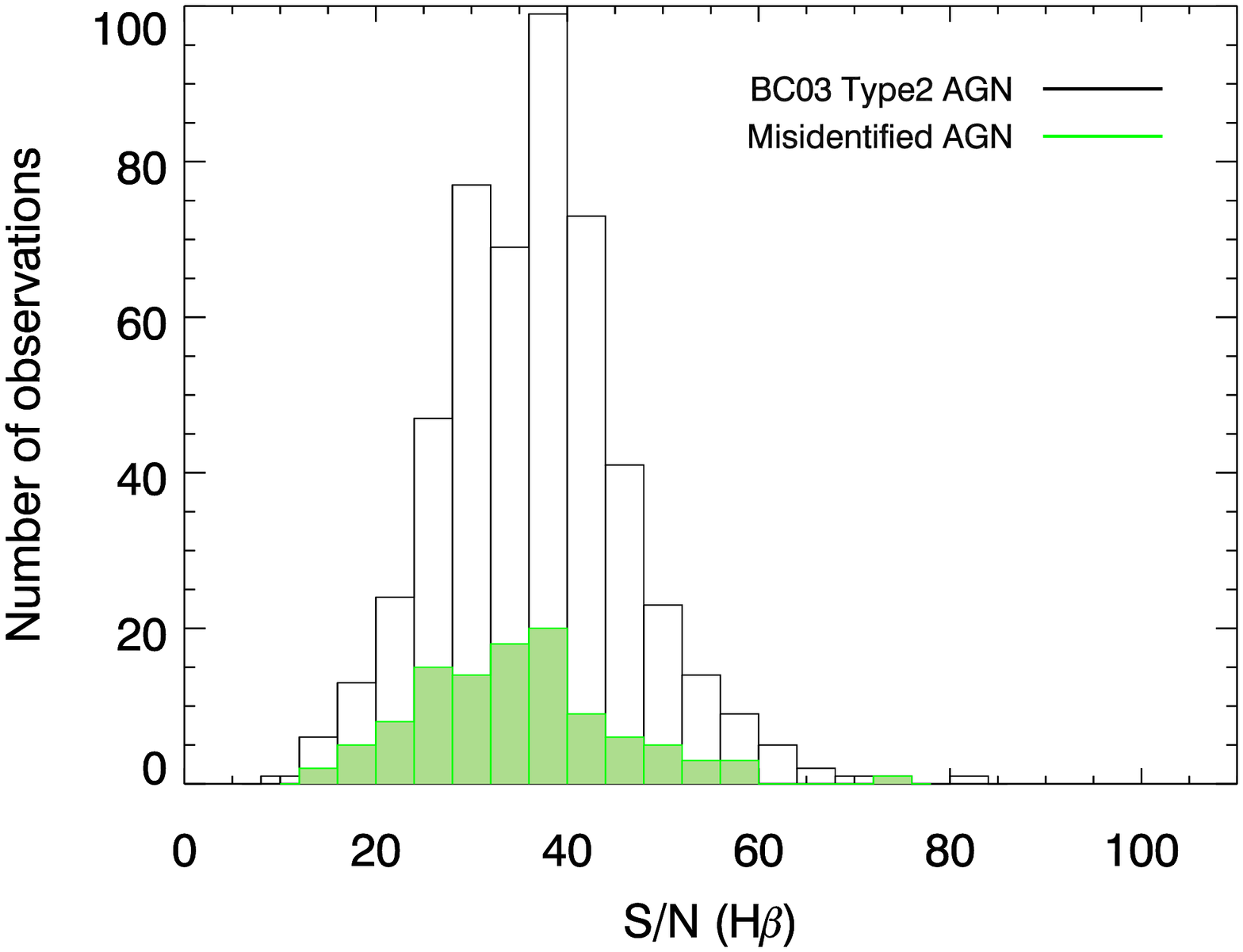} \label{fig:bcsnhba} }
\hspace*{-1.9em}
\subfloat[]{\includegraphics[scale=0.39,angle=90]{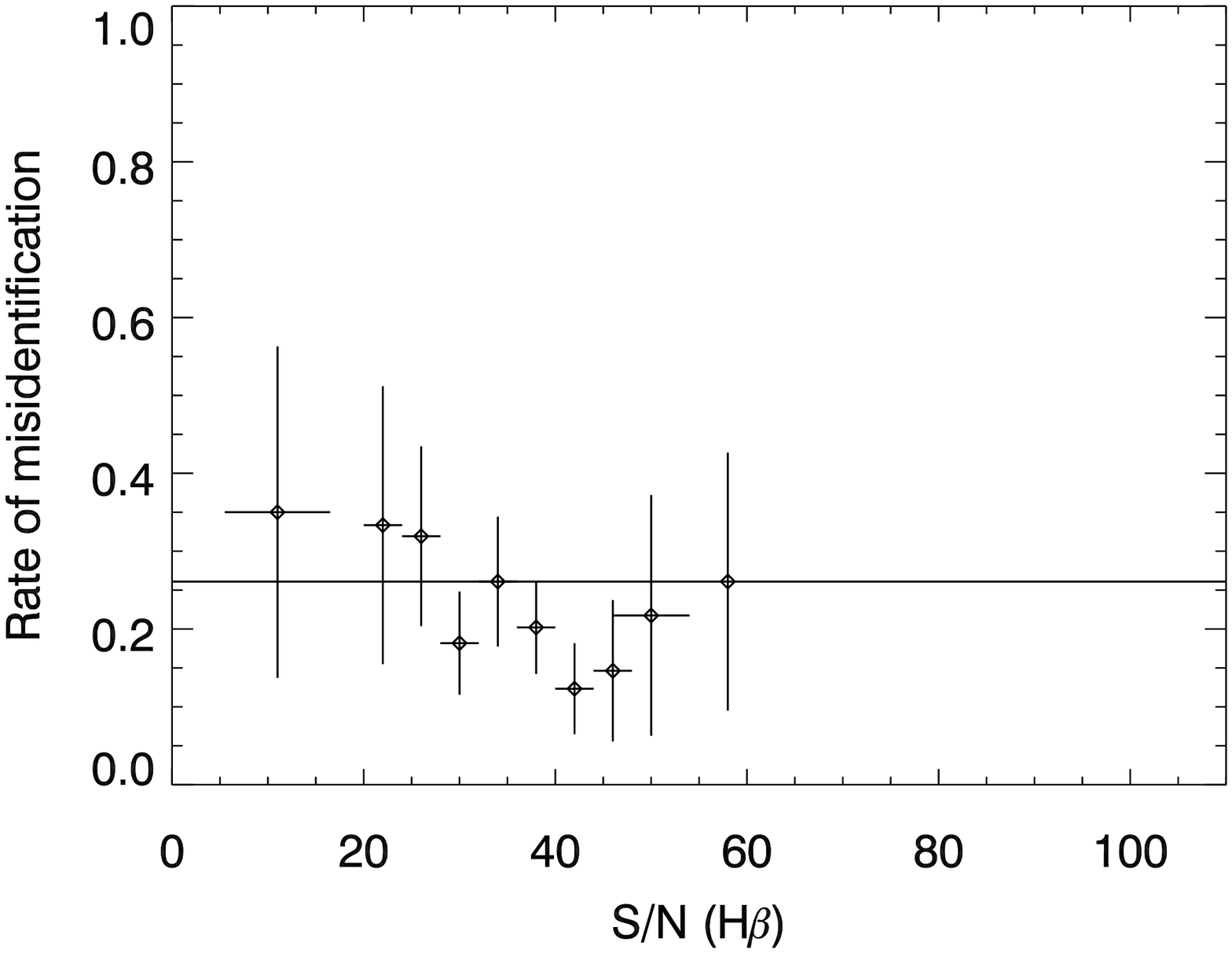}
 \label{fig:bcsnhbb}}
\\
\subfloat[]{\includegraphics[scale=0.39,angle=90]{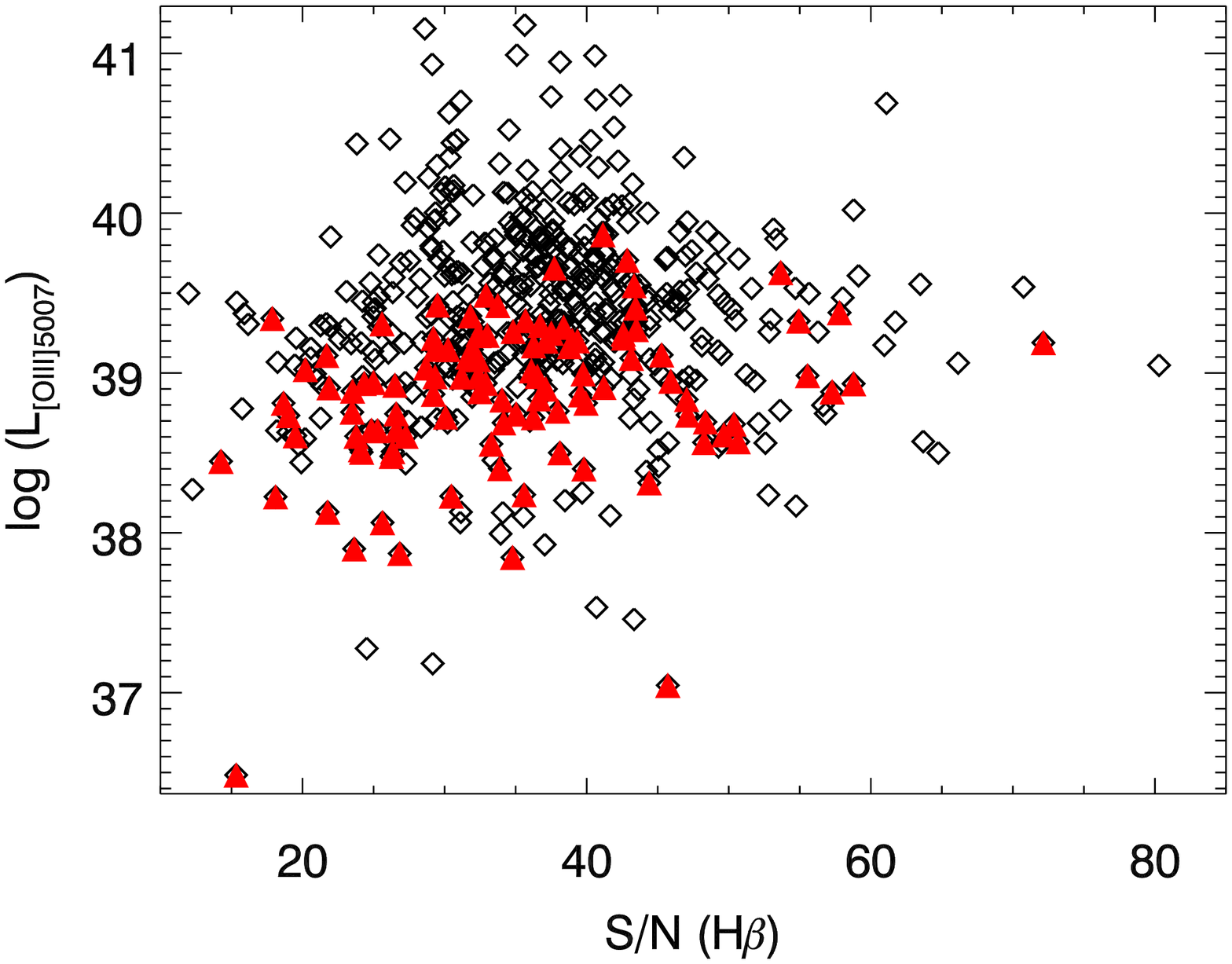}
 \label{fig:bcsnhbc}}
\caption{BC03 based misidentification as a function of data quality near  $\rm H{\beta}$ line. Panel (a): the clear histogram shows all
 type II AGN identified by using BC03 templates. The misidentified type II AGN are shown in filled green histogram. 
Panel (b): The misidentification rate as a function of continuum $\rm S/N$ near $\rm H{\beta}$ line. Panel (c): The 
luminosity of \OIII 5007 of type II AGN from BC03 template subtraction 
as a function of continuum $\rm S/N$ near $\rm H{\beta}$ line are shown as black diamonds. The misidentified type II AGN are marked by filled 
red triangles.}
\end{figure}

\begin{figure}  
\subfloat[]{\includegraphics[scale=0.39,angle=90]
    {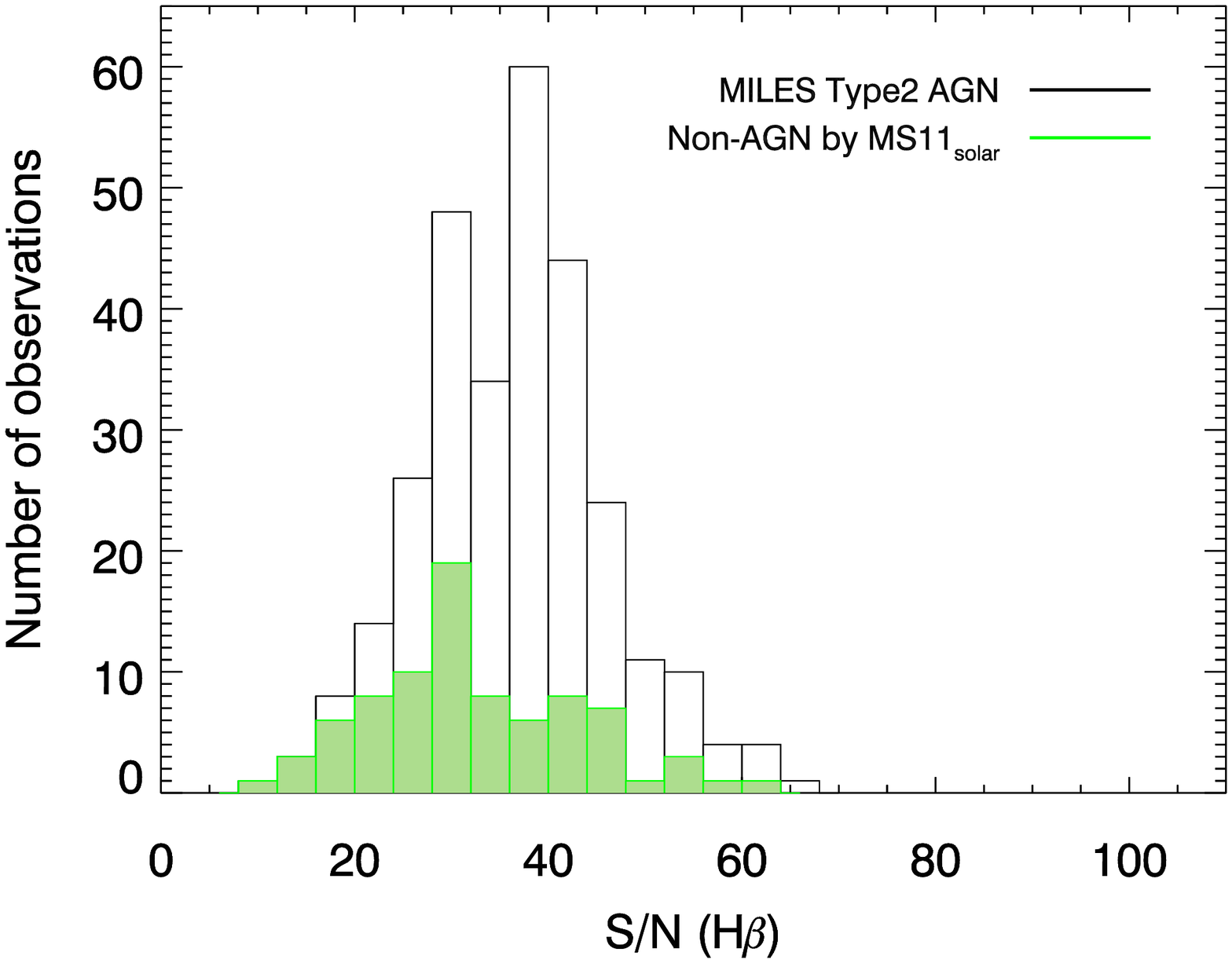}  \label{fig:portssnhba} }
\hspace*{-1.9em}
\subfloat[]{\includegraphics[scale=0.39,angle=90]{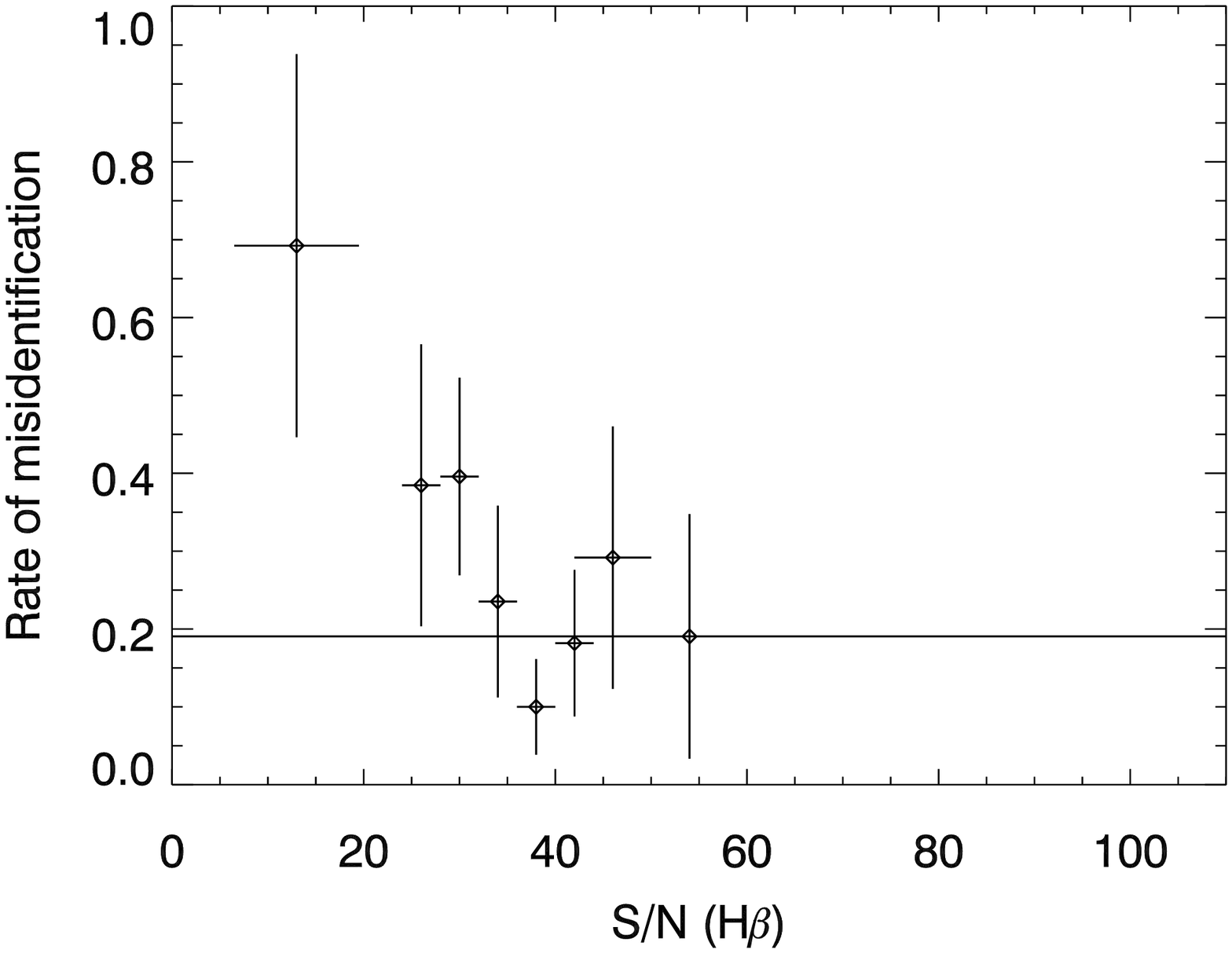}\label{fig:portssnhbb}}
\\
\subfloat[]{\includegraphics[scale=0.39,angle=90]{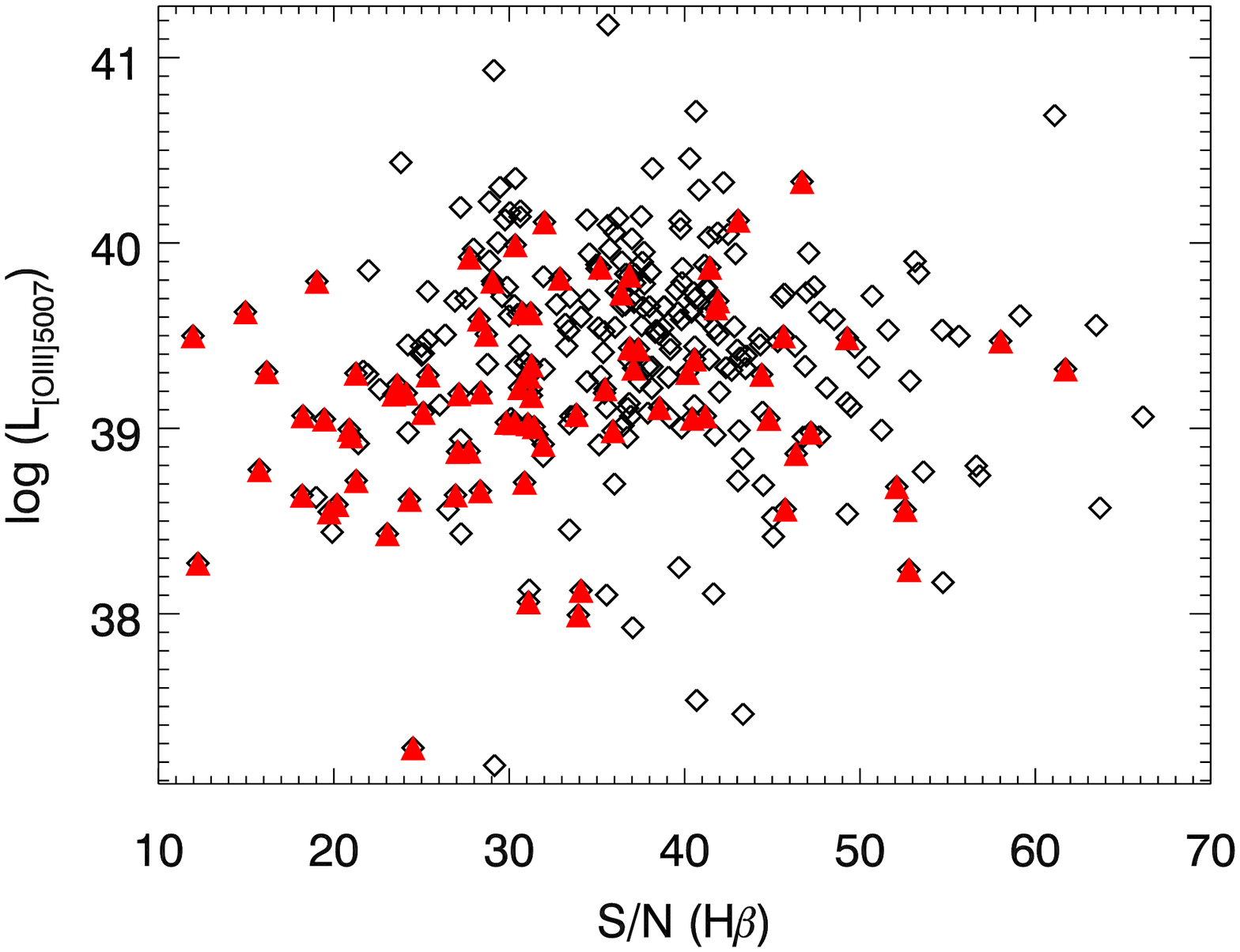} \label{fig:portssnhbc}}
\caption{MS11$_{solar}$ based misidentification as a function of data quality near  $\rm H{\beta}$ line. 
Panel (a): the clear histogram shows all
type II AGN identified by using MS11$_{solar}$ templates. The misidentified type II AGN are shown in filled green histogram. 
Panel (b): The misidentification rate as a function of continuum $\rm S/N$ near $\rm H{\beta}$ line. Panel (c): The 
luminosity of \OIII 5007 of type II AGN from MS11$_{solar}$ template subtraction 
as a function of continuum $\rm S/N$ near $\rm H{\beta}$ line are shown as black diamonds. The misidentified type II AGN are marked by filled 
red triangles.}
\label{fig:portssnhball}
\end{figure}

We show the result from MS11$_{solar}$ templates in Figure~\ref{fig:portssnhball}.
The distribution of S/N($\rm {H}{\beta}$) for type II AGN identified from MILES templates, but not from
MS11$_{solar}$ templates, are shown in the green histogram. Misidentification 
rate as a function of S/N($\rm {H}{\beta}$) is shown in Fig~\ref{fig:portssnhbb}. The result is similar 
to the comparison between MILES and BC03, except for
the first S/N bin that contains low statistics.
We also explore the dependence between \OIII5007 luminosity and S/N($\rm {H}{\beta}$). As shown
in Fig~\ref{fig:portssnhbc}, misidentified type II AGN are shown as filled red triangles. No strong correlation 
is found. The misidentification of type II AGN is, therefore, not due to data quality.

\section{Effects of young stellar populations}\label{appyoungssp}

\subsection{Young stellar populations}

Comparing parameters of the BC03 and MILES models in Table~\ref{Sspmodels}, 
we note that BC03 contains younger stellar populations than MILES. 
Young stellar populations are hard to model due to the lack of empirical observations. 
BC03 models have young stellar populations with a corrected continuum but their
 lines have not been corrected \citep[][Section 2.2.3]{BC03}, especially at non-solar metallicities \citep{G05}.
We instead resort to theoretical young stellar populations.
We expand the MILES model by adding young theoretical stellar population templates from
G05 to make up for the fact that there are not many empirical stellar libraries covering that parameter space.
As pointed out by \citet{CharlotFall00}, stellar populations younger than $\sim$ 3-4 Myr
do not contribute to the observed spectra, because their absorption features are hidden behind their 
optically thick H{\sc ii} cloud. We therefore have confidence that
using the models by \citet{G05} at a youngest age of 4 Myr is adequate.
To check the consistency between the G05 theoretical model and MILES model, we
compare the spectra of their populations at the common age of 63 Myr, the youngest stellar population
 in the MILES model,
at metallicities of $\rm[Fe/H]=-0.4$ and $\rm[Fe/H]=0.0$ (solar metallicity). 
As shown in Fig~\ref{fig:milesvsg05}, 
the G05 and MILES models are in general consistent with each other even at the boundary of the MILES model,
with less than 10 percent deviation in their residual spectra. 

\begin{figure}[h]
\centering
\includegraphics[scale=0.6]{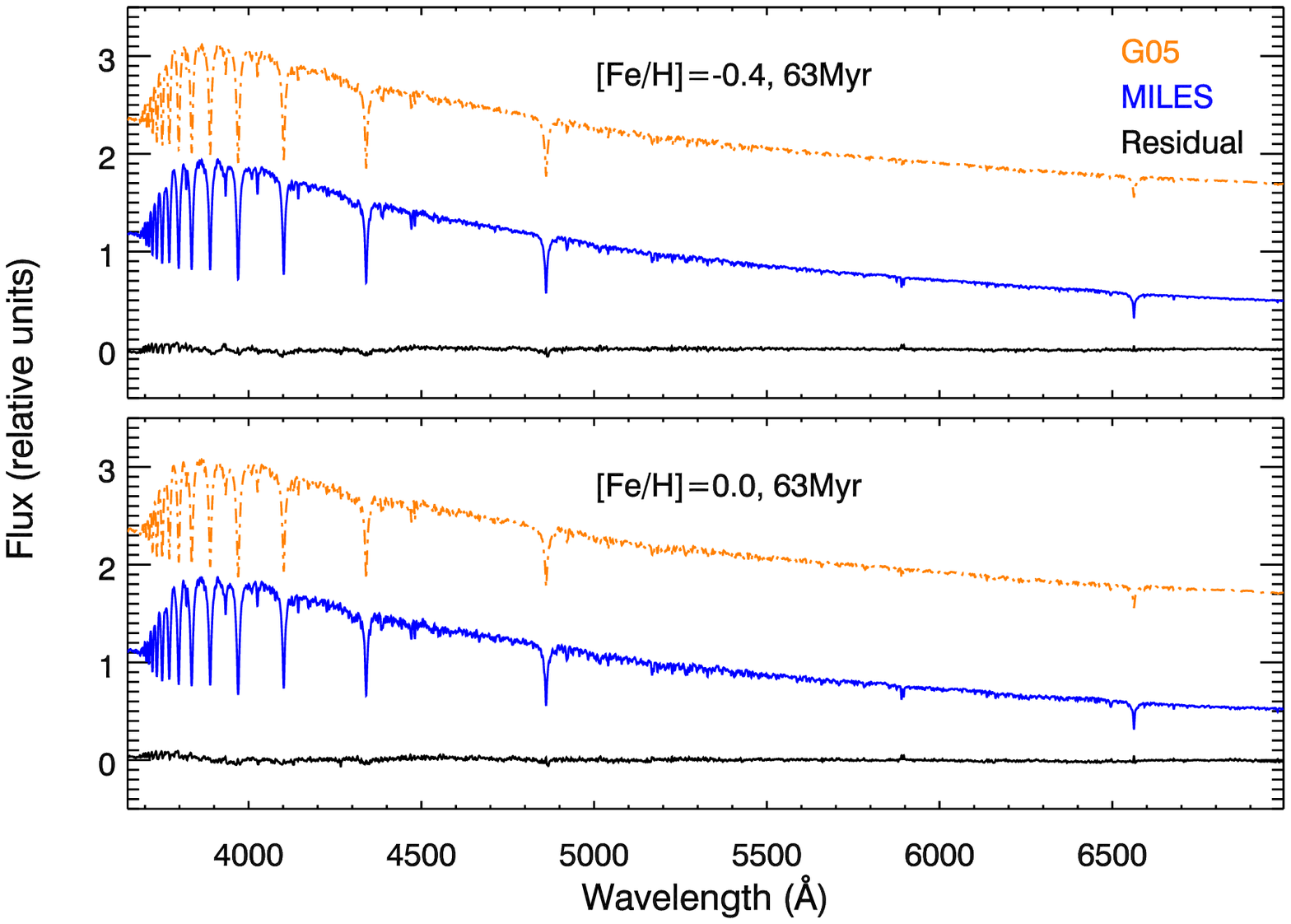}
\caption{G05 templates of 63 Myr at two metalicities compared with the MILES model. G05 templates were
smoothed to the same resolution as MILES library of FWHM=2.5 \AA. The residual spectra are the difference 
between these two sets of models.}
\label{fig:milesvsg05}
\end{figure}

\subsection{AGN identification with the addition of young stellar populations}
The consistency discussed above gives us confidence that mixing theoretical and 
empirical models does not introduce systematic effects.  Because the wavelength range in \citet{G05} is slightly 
shorter than MILES, we truncate the MILES model to the common
 wavelength range $\lambda\lambda 3500-7000$ \AA.  All stellar population models younger than 
63 Myr from \citet{G05} were broadened to the same resolution as MILES.

\begin{figure}
\centering
\includegraphics[scale=0.6]{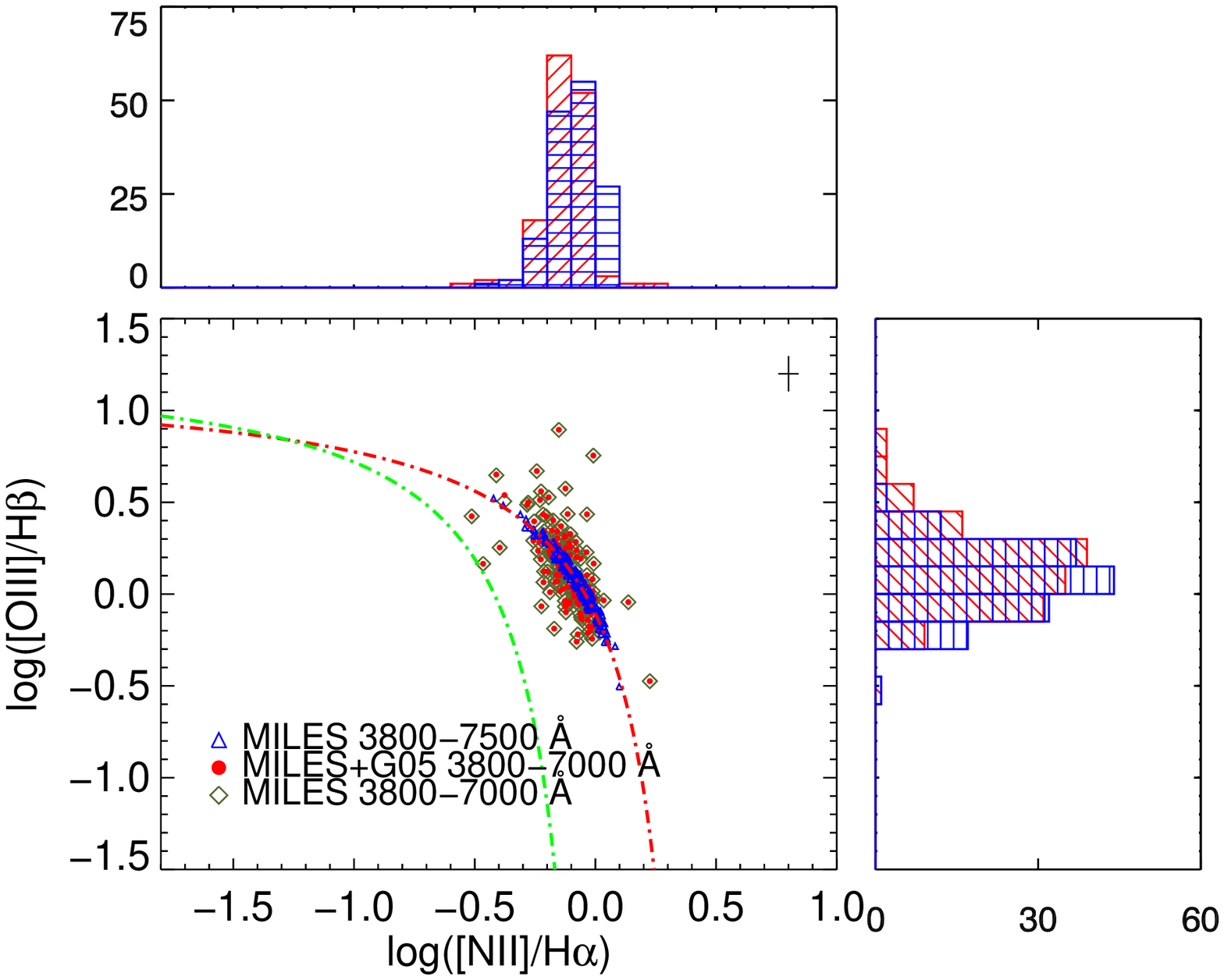} 
\caption{The same galaxies based on MILES template subtraction within 0.02 dex around the \citet{K01} boundary 
selected as in Fig~\ref{fig:BPTth13miles}, with addition of the line ratios derived from 
young-stellar- population-extended MILES templates (red dots) and wavelength-truncated MILES templates 
(olive diamonds). Agreement of red dots and olive diamonds shows that adding young stellar population 
templates does not change the line ratios. The line ratios derived from truncated templates cause a larger
 scatter around the \cite{K01} boundary.  The distributions of line ratios are histogramed
 on the side of the axes. The colors of histograms are the same as 
for the BPT diagram.}
\label{fig:BPTbcmilesg05}
\end{figure}

Following our initial strategy, we limit the sample within 0.02 dex around
 the \citet{K01} boundary in BPT diagram to investigate the line ratio variations.
 We fit the SDSS spectra using the young-population-extended MILES templates.
 We found that only three objects out of the 145 boundary galaxies contain young population components, with
 a contribution less than four percent in each of the cases. Fits for the other 142 galaxies 
 give almost exactly the same result as using only MILES templates in the same wavelength coverage, as shown in 
 Fig~\ref{fig:BPTbcmilesg05}. In fact, even those three objects show only small shifts in the line ratios, 
 $\sim$0.03 in \OIII$/ \rm H{\beta}$ and $\sim$0.002 in  \NII$/ \rm H{\alpha}$, far smaller than the typical line 
 ratio errors. As shown by the olive diamonds (MILES $+$ G05 young populations) overlapped with the red dots 
 (MILES only, wavelength truncated to the same range as MILES $+$ G05), there are no significant changes 
 resulting from adding younger templates.

\section{Wavelength range dependence}\label{appwavelength}

We also considered whether the wavelength range used for the 
spectrum fitting and absorption and continuum component subtraction may play a role in AGN identification.  
We performed this test using both the BC03 templates and MILES templates.

\begin{figure}
\subfloat[]{\includegraphics[scale=0.37,angle=90]
    {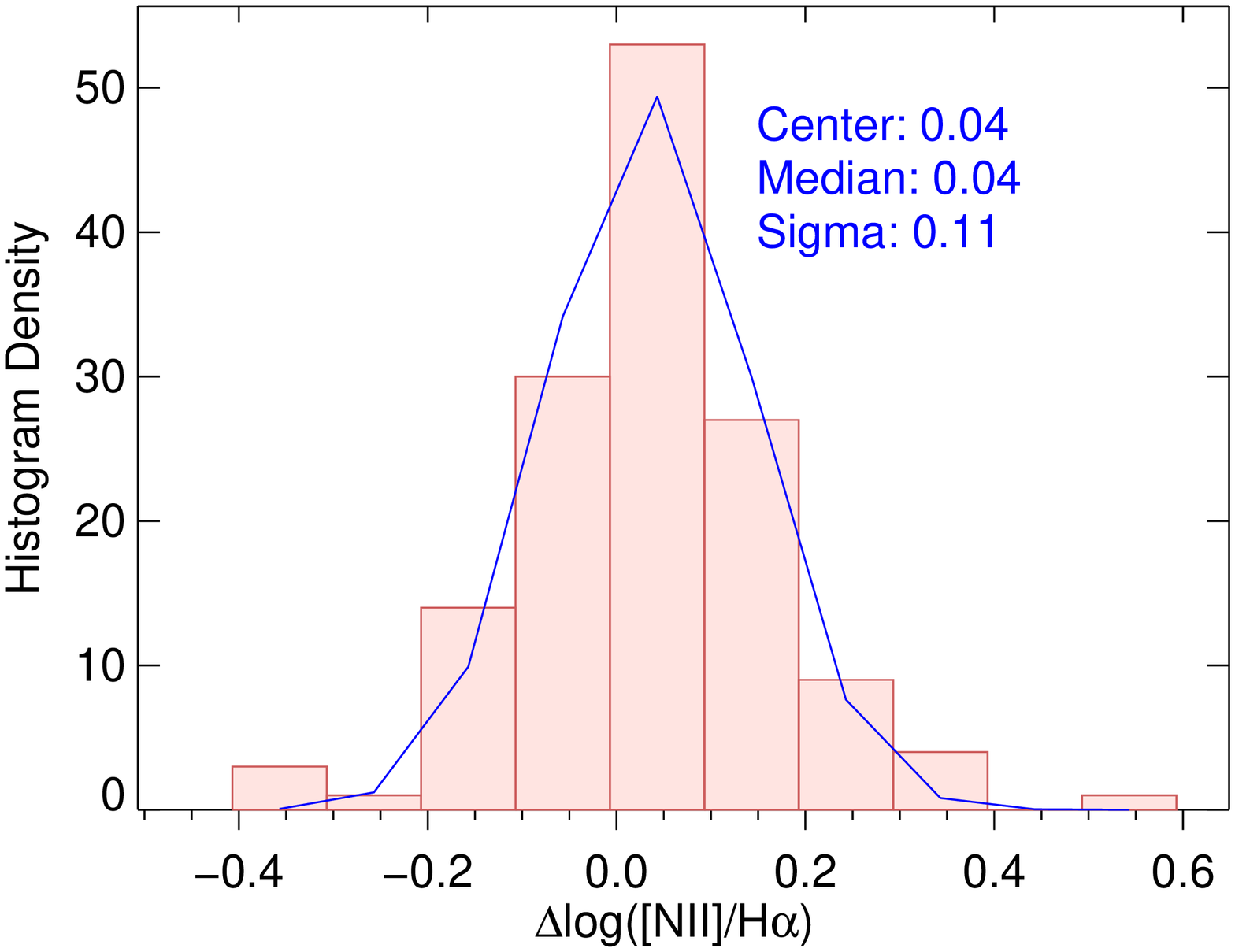}} 
\subfloat[]{\includegraphics[scale=0.37,angle=90]{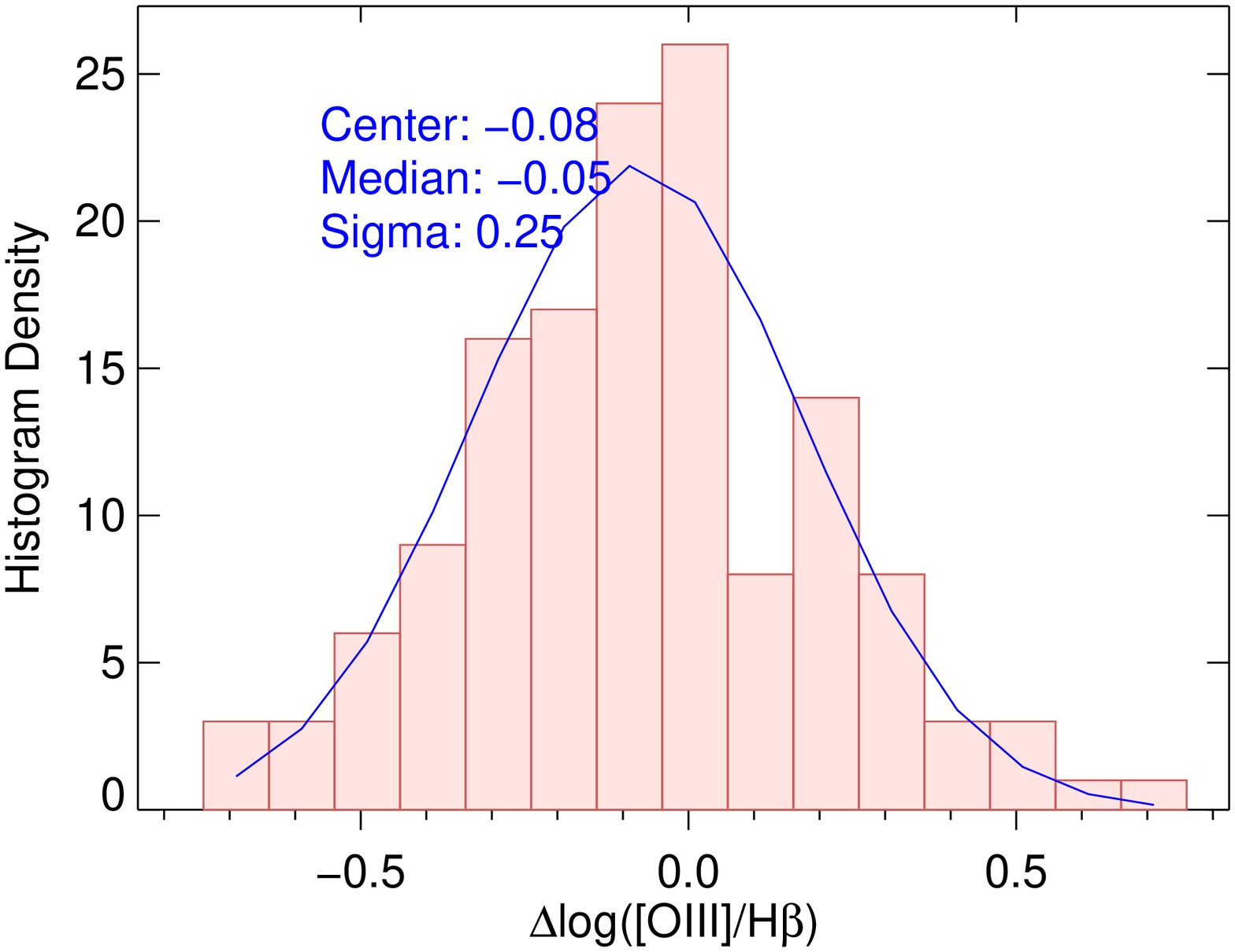}}
\caption{Distribution of line ratio differences between the fits with MILES model in
wavelength range of 3800--7500\AA\ and 3800--7000\AA. Left panel shows the difference in the 
\NII/$\rm H{\alpha}$ ratio; right panel shows that difference in the \OIII/$\rm H{\beta}$ ratio.}
\label{fig:milestesthist}
\end{figure}

As mentioned above, we expanded the MILES models with young stellar populations with G05 with truncated 
wavelength 3500--7000 \AA\ since the red wavelength limit of G05 models is 7000\AA. 
The line ratio variation from wavelength truncation of MILES templates
is shown in Fig~\ref{fig:BPTbcmilesg05}. Compared to the line ratios derived from 
 the full wavelength range of MILES, $\lambda\lambda 3800-7500$\AA, 
 the result with truncated MILES templates shows scattered line ratios around the \citet{K01} boundary
but no strong systematic effects.
 Fig~\ref{fig:BPTbcmilesg05} illustrates the distribution of the line ratios: by truncating 
only 400--500 \AA\ in the red, the line ratios scatter more around the \citet{K01} boundary, sharing the
similar ranges of \OIII/$\rm H{\beta}$ and \NII/$\rm H{\alpha}$. We compare the line ratio differences in 
Fig~\ref{fig:milestesthist}. It shows that the centre of \NII/$\rm H{\alpha}$ is shifted by 0.04 dex, and the 
centre of \OIII/$\rm H{\beta}$ is shifted by -0.05 dex. The line ratio offsets of both \NII/$\rm H{\alpha}$ and 
\OIII/$\rm H{\beta}$ are consistent with zero within the errors.

\begin{figure}
\centering
\includegraphics[scale=0.6,angle=90]{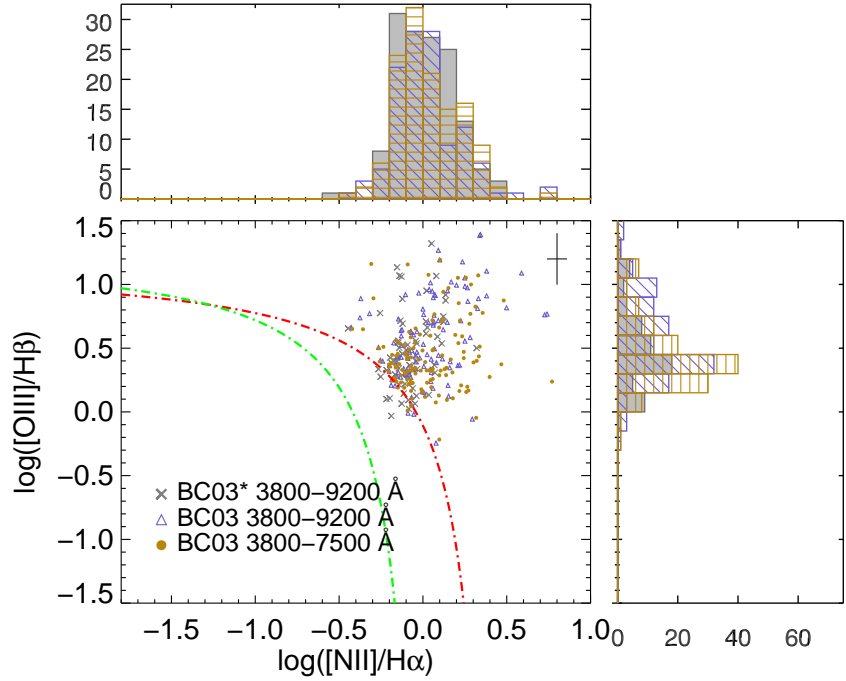} 
\caption{BC03-based line ratios for 145 objects near the \citet{K01} boundary as defined in Fig~\ref{fig:BPTth13miles}. 
The grey crosses are directly from SDSS DR8 archive, the purple triangles are from this work, 
the golden dots are from the wavelength-truncated BC03 templates.
The distributions of line ratios are histogramed on the side of the axes. The colors are the same as the line ratios. }
\label{fig:bcall}
\end{figure}

We changed the fitting wavelength
range to be $\lambda\lambda 3800-7500$ \AA\ in testing the BC03 templates. The results are shown in
Fig~\ref{fig:bcall} to illustrate the difference from the original fitting wavelength of 
$\lambda\lambda 3800-9200$ \AA.
The grey histograms show
 the distribution of SDSS DR8 line ratios, 
and the golden  histograms show the line ratios derived by truncated spectrum-fitting using BC03.
  The line ratios from truncated wavelength fit, 
 still scatter around the same region in the BPT diagram as the result from wider wavelength range.

\begin{figure}
\subfloat[]{\includegraphics[scale=0.37,angle=90]
    {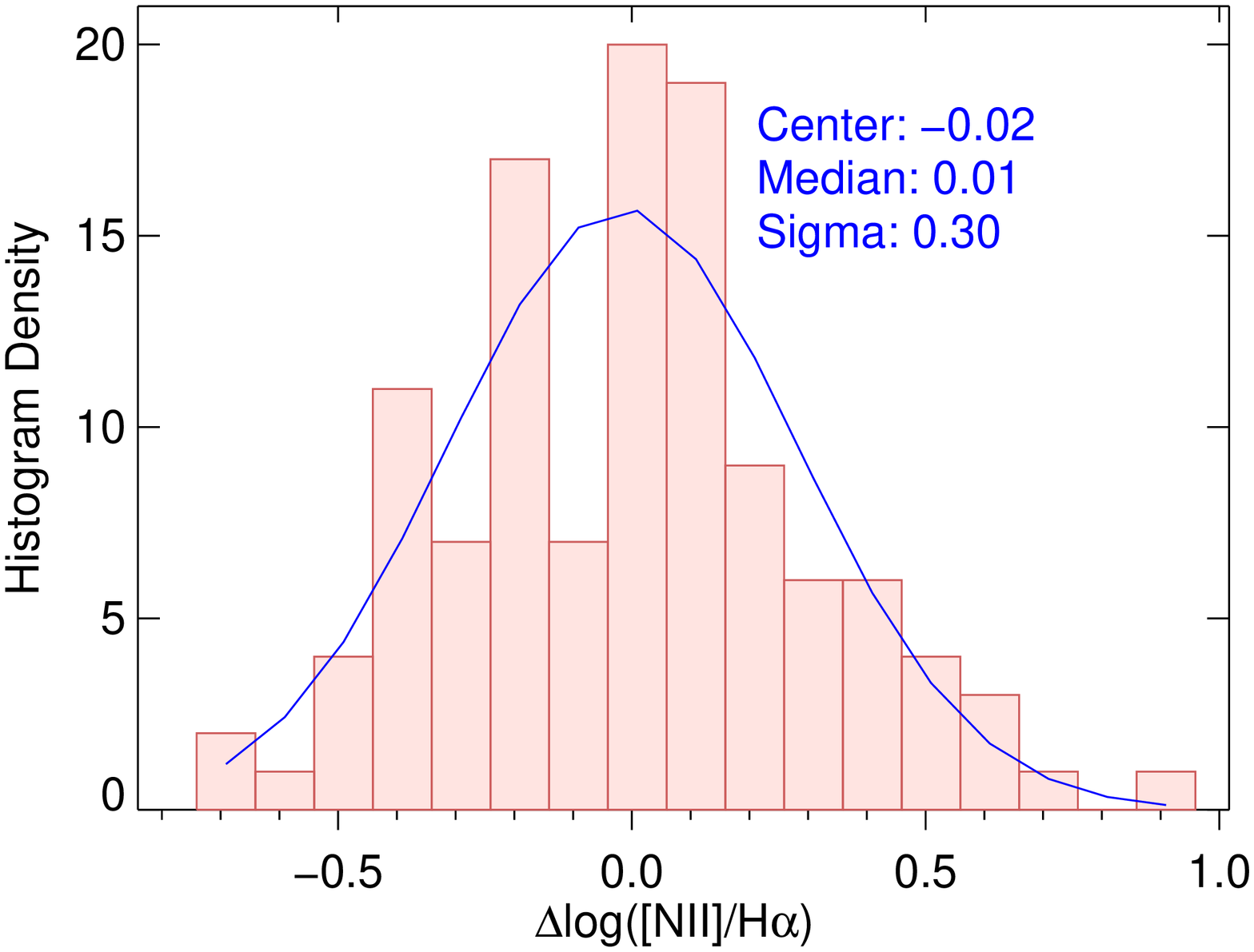}} 
\subfloat[]{\includegraphics[scale=0.37,angle=90]{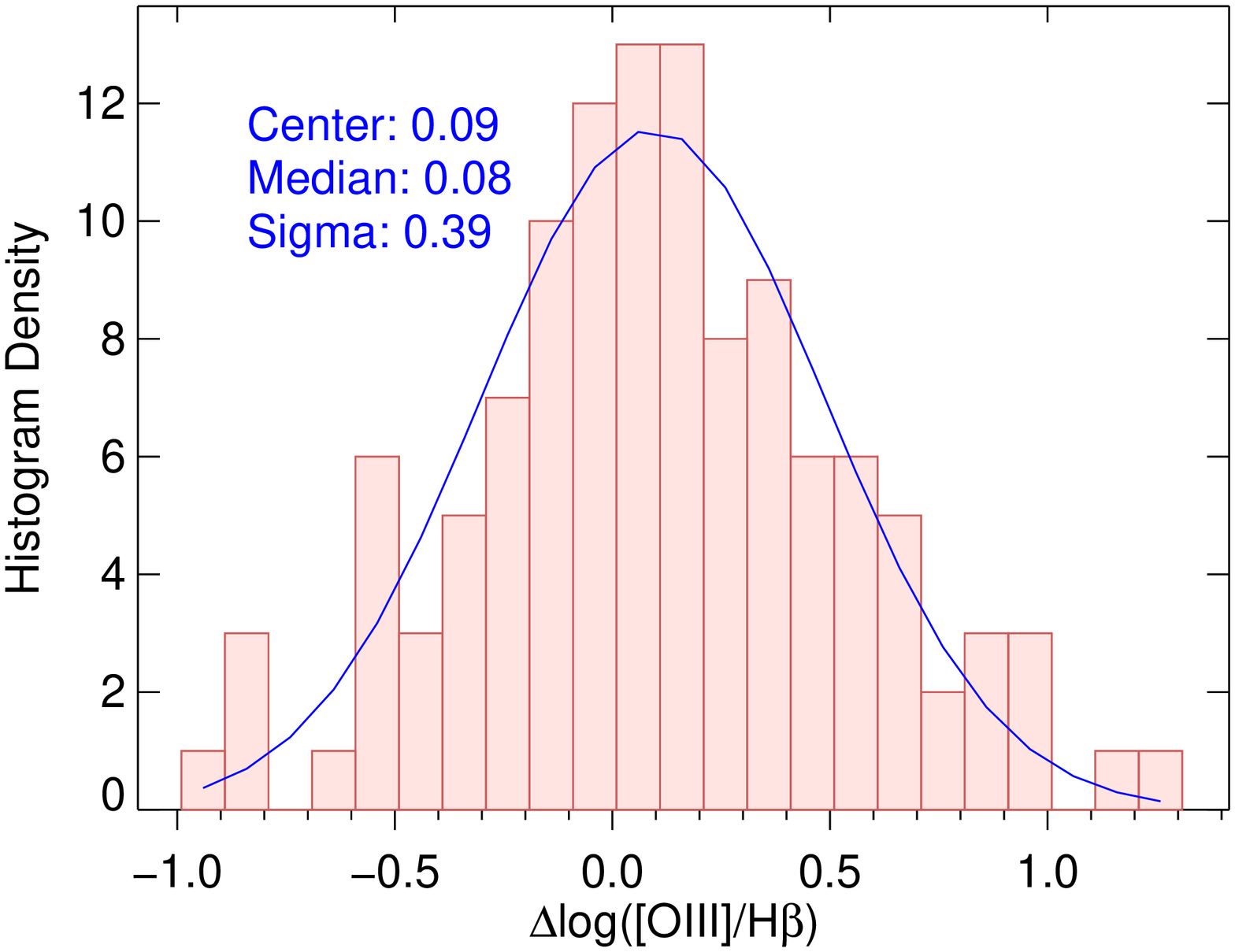}}
\caption{Distribution of line ratio differences between the fits with BC03 templates in
wavelength range of 3800--9200\AA\ and 3800--7400\AA. Left panel shows the difference of 
line ratio \NII/$\rm H{\alpha}$; right panel shows that difference of line ratio \OIII/$\rm H{\beta}$.}
\label{fig:bc03testhist}
\end{figure}

A detailed systematic analysis on the line ratios of truncated BC03 templates is shown in in Fig~\ref{fig:bc03testhist}. 
A shift of 0.08 dex is observed in \OIII/$\rm H{\beta}$, and a shift of $\sim$0.01 dex is observed in  \NII/$\rm H{\alpha}$. 
Again, the line ratio offsets of both \NII/$\rm H{\alpha}$ and 
\OIII/$\rm H{\beta}$ are consistent with zero within the errors.

\end{document}